\documentclass[10pt,journal,twocolumn,twoside]{IEEEtran}
\usepackage{graphicx}
\usepackage{epstopdf}
\usepackage{float}
\usepackage{algorithmic}
\usepackage{array}
\usepackage{amsmath}
\usepackage{amssymb}
\usepackage{mdwmath}
\usepackage{mdwtab}
\usepackage{eqparbox}
\usepackage{stfloats}
\usepackage{fixltx2e}
\usepackage{cases} 

\usepackage{xcolor}
\usepackage{makecell}

\usepackage[boxed,ruled,commentsnumbered]{algorithm2e}
\usepackage{url}
\usepackage{cite}
\ifCLASSOPTIONcompsoc
\usepackage[caption=false,font=normalsize,labelfont=sf,textfont=sf]{subfig}
\else
\usepackage[caption=false,font=footnotesize]{subfig}
\fi
\allowdisplaybreaks[4]

\newtheorem{proposition}{Proposition}

\makeatletter
\renewcommand*{\@opargbegintheorem}[3]{\trivlist
      \item[\hskip \labelsep{\bfseries #1\ #2}] \textbf{(#3):}\ }
\makeatother     

\begin{document}

\title{Robust Beamforming and Rate-Splitting Design 
for Next Generation Ultra-Reliable and Low-Latency Communications  }

\author{\IEEEauthorblockN{Tiantian~Li,
                     Haixia~Zhang,~\IEEEmembership{Senior~Member,~IEEE,}
                     Shuaishuai~Guo,~\IEEEmembership{Member,~IEEE,}
                     Dongfeng~Yuan,~\IEEEmembership{Senior~Member,~IEEE}}

\thanks{ Tiantian Li is  with the School of Information Science and Engineering, Shandong Normal University, Jinan, Shandong 250358, China, and also with the Shandong Key Laboratory of Wireless Communication Technologies, Shandong University, Jinan, Shandong 250061, China (e-mail: tiantian.li@sdnu.edu.cn.) } 
\thanks{ Haixia Zhang and Shuaishuai Guo are  with the Shandong Key Laboratory of Wireless Communication Technologies, and  the School of Control Science and Engineering, Shandong University, Jinan, Shandong 250061, China (e-mail: haixia.zhang@sdu.edu.cn; shuaishuai$\_$guo@sdu.edu.cn.)}
\thanks{ Dongfeng Yuan is with the Shandong Key Laboratory of Wireless Communication Technologies, Shandong University, Jinan, Shandong 250061, China. (e-mail: dfyuan@sdu.edu.cn.) }
}

\markboth{ }%
{Shell \MakeLowercase{\textit{et al.}}: Bare Demo of IEEEtran.cls for IEEE Journals}
\maketitle
\IEEEpeerreviewmaketitle

\begin{abstract} 
The next generation ultra-reliable and low-latency communications (xURLLC) need novel design to  provide satisfactory services to the emerging mission-critical  applications.  To improve the  spectrum  efficiency and enhance the robustness of xURLLC, this paper proposes a robust beamforming and  rate-splitting design in the finite blocklength (FBL) regime for   downlink multi-user multi-antenna xURLLC systems.  In the design, adaptive rate-splitting  is introduced  to flexibly handle the complex inter-user interference and thus improve the  spectrum  efficiency.    Taking the imperfection of the channel state information  at the transmitter (CSIT) into consideration,  a max-min user rate problem is formulated to optimize the  common and private beamforming vectors and the rate-splitting vector under the premise of ensuring the requirements of transmission latency and  reliability  of all  the users. The optimization  problem is intractable due to the non-convexity of the constraint set and the infinite constraints caused by CSIT uncertainties. To solve it, we   convert the infinite constraints into finite ones by the S-Procedure method  and transform the original problem into a difference of convex (DC) programming. A constrained concave convex procedure (CCCP) and the Gaussian randomization based iterative algorithm is proposed to obtain a local minimum. Simulation results confirm the convergence,  robustness and effectiveness of the proposed  robust beamforming and rate-splitting design in the FBL regime. It is also shown that the proposed robust design  achieves considerable performance gain in the worst user rate compared with existing transmission schemes under various  blocklength and block error rate requirements.

\end{abstract}

\begin{IEEEkeywords} Robust beamforming, rate-splitting,  finite blocklength (FBL), next generation ultra-reliable and low-latency communications (xURLLC).
\end{IEEEkeywords}


\section{Introduction}
\IEEEPARstart{U}{ltra}-reliable and low-latency communications (URLLC) have become an essential part  of  the  fifth-generation (5G)  mobile communication systems to facilitate many  mission-critical  applications such as industrial automation, intelligent transportation, telemedicine, Tactile Internet  and extended reality \cite{Enabling2019Sutton,short2019Shirvanimoghaddam,resource2021Nasir}. In  URLLC   systems, the communication is required to have no less than 99.999$\%$ transmission reliability or no more than  $10^{-5}$ block error rate (BLER), no longer than  1 ms end-to-end latency, and short transmission packet \cite{transmission2020xu,beamforming2021he}.   Nevertheless, 5G URLLC does not fulfill all the Key Performance Indicators  of diverse mission-critical applications. To provide new satisfactory services to the emerging mission-critical  applications, the next generation URLLC (xURLLC) for the sixth-generation (6G)  communication systems need novel design to  provide   higher spectrum efficiency under the premise of guaranteeing high transmission reliability and low end-to-end latency. Providing such high spectrum efficiency xURLLC services is  challenging especially  for the multi-user  communication  networks due to the existence of  complex inter-user interference.

In  the physical layer,  over-the-air transmission latency is the key component of the communication latency. In traditional data-rate-oriented networks,  transmission latency  is relatively long (say 20-30 ms) and the packet size is relatively large (say 1500 bytes), Shannon's  capacity served as a tight upper bound of the achievable data rate due to the law of large numbers. But in (x)URLLC, to reduce  over-the-air transmission latency, short packets with finite  blocklength (FBL) such as 20 bytes,  32 bytes, should be  adopted \cite{joint2020ren}.  That is to say, Shannon's achievable capacity formula based on the infinite blocklength (IFBL) coding   theory  is no longer applicable \cite{Toward2016Durisi}. In the FBL regime of (x)URLLC, the  blocklength of transmission data is shortened and thus the transmission latency is reduced. This makes the BLER not negligible \cite{Wirelessly2017Khan}. To investigate the relationship between the achievable rate, the  blocklength and the BLER, Polyanskiy et al. approximated the theoretical boundary of achievable rate  with FBL  in 2010  and found that it was a complicated function of  transmitting power, the blocklength and the BLER \cite{channel2010Polyanskiy,Quasi2014Yang}. In 2015, Yang and Polyanskiy et al. studied  the maximum channel coding rate of quasi-static fading channel and additive Gaussian white noise channel under the constraints of BLER,  blocklength and transmission power \cite{Optimum2015yang}. Furthermore, Durisi and Polyanskiy  et al.  investigated the maximum achievable  rate for a given blocklength and BLER in a quasi-static multi-antenna Rayleigh fading channel \cite{Quasi2014Yang}, and analyzed the trade-off among transmission latency, reliability and achievable rate \cite{Short2016Durisi}.

Based on the above  works, further investigation on multi-user multi-antenna communications in the FBL regime has attracted extensive researches. In the precoding or beamforming   design with FBL, traditional multi-user linear precoding  is widely considered, in which each message is encoded into an independent data stream and then all the encoded streams are spatially multiplexed through linear precoding  at the transmitter.  At each receiver, the expected signal is detected by treating all the inter-user  interference  as noise. This scheme can exploit  the spatial multiplexing gain to help improve the  system performance when the spatial degrees of freedom (DoF) is relatively high.  As the asymptotic capacity based on the FBL coding theory is a complex function of  the beamforming vectors, transmission power, transmission block and BLER,  the corresponding beamforming  problem in the FBL regime is highly non-convex and computationally challenging \cite{energy2021Singh,resource2021Nasir}. To deal with it, \cite{energy2021Singh} investigated the energy efficiency maximization problem by optimizing the beamforming vectors  and the BLER in the  FBL  regime for downlink multi-user multiple-input single-output (MISO) networks. The non-convex  problem was accurately approximated by applying the Dinkelbach method and by approximating the channel dispersion in the high signal-to-interference-plus-noise ratio (SINR) regime and the entire  SINR regime, respectively.  Considering  user fairness,  \cite{resource2021Nasir} studied the beamforming  problem aiming at maximizing the users' minimum rate in downlink MISO URLLC. The  path-following algorithms were developed to solve the non-convex problem, and could converge to a locally optimal solution. Further,  \cite{beamforming2021he} analyzed the function of achievable rate with FBL and some important insights had been discovered such as the analytical solution of the minimum rate requirement with respective to the SINR. The authors also proposed effective algorithms to optimize the beamforming vectors  and power allocation  to obtain a local optimum solution with low computational complexity.

In multi-user MISO communication systems, if the DoF is limited to the antenna number or the correlation of user channels, the inter-user interference is too strong  to suppress, leading to low spectrum efficiency. In this case,  the above traditional transmission scheme   fails to perform well. Fortunately, a generalized and unifying interference management strategy called rate-splitting  has been proposed and extensively studied in traditional data-rate-oriented communication networks  \cite{ Clerckx2016rate,mao2018rate,Clerckx2019rate}. Different from the above mentioned non-rate-splitting transmission scheme (namely NoRS scheme) which fully treats any interference as noise, the rate-splitting strategy is to decode part of the interference and treat the remaining part as noise. In specific,  at the transmitter the  messages  for all the users are first split and then encoded into one common stream which is decoded by all the receivers, and multiple private streams which are decoded by the corresponding receivers. At each receiver, the superimposed common and private streams are sequentially decoded  using  successive interference cancellation (SIC) technique.  As expected, the rate-splitting strategy  can  cope with a wide range of propagation conditions and achieve a significant spectrum /energy efficiency improvement in different network loads \cite{joundeh2017rate,Mao2019rate,Yalcin2020rate,chen2020joint},   user deployments \cite{mao2018energy,full2022li}, and  channel state information  at the transmitter (CSIT)  inaccuracy \cite{Joudeh2016sumrate,joudeh2016robust,Yin2020rate}. Motivated by the significant performance gain of rate-splitting in the IFBL regime, Clerckx et al. introduced the rate-splitting design to URLLC systems to maximize the weighted sum rate or  the Max-Min Fairness (MMF) rate \cite{xu2021rate,dizdar2021rate,xu2022max}. In these works, it is assumed  that the  CSIT can be perfectly acquired.
 
However, in various practical (x)URLLC systems, such as the communication networks equipped with multiple  antennas, and the communication networks where vehicles move in high speed, the CSIT is difficult to acquire accurately due to  estimation errors in Time Division Duplex (TDD) systems and quantization errors in Frequency Division Duplex (FDD) systems, or the CSIT is delayed/outdated due to  the Doppler effect. Designing   beamforming schemes using imperfect/outdated  CSIT  will introduce residual interference at the receivers,  thus dramatically damage the   system performance. For example,  the actually resulted transmission latency  will increase and then violate the latency requirements of the users. The transmission reliability will have a certain decrease,  resulting in an inability to support mission critical applications. 

To improve the  spectrum efficiency of xURLLC and  meanwhile enhance the system robustness, this paper proposes a robust beamforming and  rate-splitting design in the finite blocklength (FBL) regime for  downlink multi-user multi-antenna xURLLC systems under CSIT imperfections. The main contributions of this paper are summarized as follows.

\begin{itemize}
\item  We propose a beamforming and rate-splitting design in the FBL regime for  spectrum efficient and robust  xURLLC.  Specifically, under the premise of ensuring the requirements of transmission latency and transmission reliability  of all the users, we formulate a max-min user rate problem to jointly design the common and private beamforming vectors  and the  rate-splitting vector   over all possible channels in the uncertain region.

\item 
The optimization  problem is intractable due to the non-convexity of the constraint set and the infinite constraints caused by CSIT uncertainties. 
 To solve it, we   convert the infinite constraints into finite ones by the S-Procedure method  and transform the original problem into a difference of convex (DC) programming.

\item  We propose an efficient  iterative algorithm based on  the constrained concave convex procedure (CCCP) and  the Gaussian randomization to solve the resulted DC programming, which outputs the robust beamforming and rate-splitting  design in the FBL regime (RB-RS-FBL).  Moreover,   an  intuitive initial feasible point search algorithm is further proposed to start the iterative algorithm.

\item Extensive simulations are carried out and the simulation results  confirm the convergence, the robustness and the effectiveness of the proposed iterative algorithm. Also, it  reveals  that  the proposed {RB-RS-FBL} design achieves considerable performance gain in the worst user rate compared with existing transmission schemes under various  blocklength and BLER requirements.
\end{itemize}

The remainder of this paper is organized as follows. Section \ref{System} introduces the system model. The problem formulation and transformation are presented in Section \ref{Formulation}. The robust beamforming and rate-splitting design is proposed in \ref{Design} including the iterative algorithm and initial feasible point search algorithm. In Section \ref{Simulation}, the simulation results and analysis are given. Finally, Section \ref{conclusion} concludes the whole paper.

\emph{Notations:} Vectors and matrices are denoted by boldface lowercase and uppercase letters, respectively. $\mathbb{C}^{N \times 1}$ represents the $N$-dimensional complex vector space. $(\cdot)^T$ and $(\cdot)^\dag$ denote the transpose operator and  complex conjugate transpose operator, respectively. $\mathcal{CN} (a,B)$ represents the circularly symmetric complex Gaussian random distribution with mean a and variance B. $\mathbf{A} \succeq \mathbf{0}$ denotes a positive semidefinite matrix. $\mathbb{E} [\cdot]$ means expectation. $\bigtriangledown A (\mathbf{x}^{(i)})$ denotes the derivative of function $A (\mathbf{x})$ with respect to vector $\mathbf{x}$, evaluated at point $\mathbf{x}^{(i)}$. The absolute value of a scalar is denoted by $|\cdot|$,  the $l_2$-norm of a vector is denoted by $||\cdot||$. $\mathbf{I}_N$ denotes a $N \times N$ identity matrix. tr($\cdot$) and rank($\cdot$) denote the trace and rank operator, respectively.  $\mathbb{H}_+^N$ represents a $N$-by-$N$ Hermitian positive semidefinite matrix set.

\section{System Model}\label{System}

\begin{figure}[!t]
  \centering
  \includegraphics[width=0.49\textwidth]{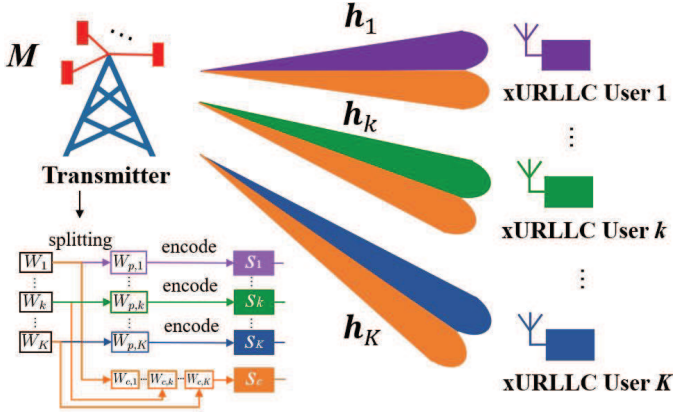}\\
  \caption{ System model of the proposed transmission  scheme for xURLLC.} 
  \label{fig_system-model}
\end{figure}
In this paper, we consider a downlink multi-user MISO system, where an $M$-antenna transmitter provides xURLLC services for $K$ single-antenna  users  indexed by $\mathcal{K}:=\{ 1,...,k,...K\}$, as shown in Fig. \ref{fig_system-model}. The transmitter sends  messages  $\{ W_1,...,W_k,...,W_K \}$ to user $1,...,k,.., K$ individually in the FBL regime. To handle the interference among users flexibly and then enhance the xURLLC system performance, rate-splitting is adopted to access all the users. At the transmitter, message $W_k$ for user $k$ is first split into a  common part $W_{c,k}$ and a private part $W_{p,k}$, $k \in \mathcal{K}$. The common parts $\{ W_{c,1},...,W_{c,k},...,W_{c,K} \}$   are combined and co-encoded into one common stream  $s_c$ using a codebook shared among all receivers. The private part $W_{p,k}$ is encoded into stream $s_k$ intended for user $k$. All the encoded  streams $s_c$ and $\{ s_1,...,s_k,...,s_K\}$ are launched with normalized power, $ \mathbb{E}[|s_{c}|^2] =...\ \mathbb{E}[|s_{k}|^2]=... \ \mathbb{E}[|s_{K}|^2]= 1$. By adopting linear transmit beamforming, the signal vector, $\mathbf{s}$, to be sent by the transmitter can be generated by superimposing all of them together with different power levels as
\begin{align} 
 \mathbf{s}= \mathbf{w}_{c}  s_{c}+\sum_{k=1}^{K} \mathbf{w}_{k}  s_{k},
\end{align} 
where $\mathbf{w}_{c} \in \mathbb{C}^{M \times 1}$ and $\mathbf{w}_{k} \in \mathbb{C}^{M \times 1}$ are the corresponding beamforming vectors for the common and private messages, respectively. Since the normalized symbol power is assumed, the transmission power of the transmitter depends on the covariance values of the beamformers,
\begin{align} \label{transmission power}
  P_t (\mathbf{w}_{c},\mathbf{w}_{k})=  \mathbf{w}_{c}^\dag \mathbf{w}_{c}+\sum_{k=1}^{K} \mathbf{w}_{k}^\dag \mathbf{w}_{k}.
\end{align} 

At  user $k  \in \mathcal{K}$, the received signal can be expressed as
\begin{align}\label{signal_CCU_N}
y_{k} = \mathbf{h}_{k}^\dag   \mathbf{w}_{c}  s_{c} 
+\mathbf{h}_{k}^\dag   \mathbf{w}_{k}  s_{k} 
+ \sum_{j=1,j \neq k}^{K} \mathbf{h}_{k}^\dag   \mathbf{w}_{j}  s_{j}  +n_k ,\
 \forall k \in \mathcal{K},
\end{align} 
where $\mathbf{h}_{k}^\dag \in \mathbb{C}^{1 \times M}$ denotes $1 \times M$ complex-valued channel coefficient vector from the transmitter to user $k$, $n_k$ represents the additive white Gaussian noise (AWGN), $ n_k \sim \mathcal{CN} (0,\sigma_{k}^{2})$. Note that in this work a quasi-static flat  fading model is considered for all the channels, i.e., they would keep constants over one time block which is smaller than the coherence time  \cite{xu2019robust} but vary independently among different time blocks.

At  receiver $k$, SIC is applied to detect  information of the common stream $s_c$ and the private stream $s_k$, sequentially. To be specific, user $k$ first detects $s_c$ by treating  all the private streams $\{ s_1,...,s_k,...,s_K\}$ as noise. According to the  asymptotic capacity based on the FBL coding theory, the  achievable common data rate in unit of nats/s/Hz to decode $s_c$ at user $k$ can be approximated  by  \cite{resource2021Nasir,beamforming2021he}
\begin{align}
&R_{c,k}(\mathbf{w}_c,\mathbf{w}_k,\mathbf{h}_k )=  \ \mathrm{ln} (1+ { \mathbf{\xi}_{c,k} } )  -
 \sqrt{\Lambda( \mathbf{\xi}_{c,k} ) }\ \frac{ Q^{-1}(\epsilon_k )}{\sqrt{L_{c,k}} } ,\notag \\
& \quad  \quad    \ \quad  \quad    \quad \quad  \quad    \quad   \quad  \quad  \quad  \quad    \quad \quad \quad \quad  \quad    \quad   \forall k\in \mathcal{K},
\end{align}
where $L_{c,k}$ is the  blocklength of stream $s_c$, $\epsilon_k$ is the block error rate, $  \Lambda(\mathbf{\xi}_{c,k})= (1-(1+ \mathbf{\xi}_{c,k} )^{-2}) $ denotes the channel dispersion and  
$ Q^{-1} (\cdot) $ denotes the inverse of the Gaussian Q-function $ Q (x)=\int_{x}^{\infty} 1/\sqrt{2\pi} \ e^{\frac{-t^2}{2}} dt$. 
${\mathbf{\xi}_{c,k}}$ denotes the signal-to-noise-ratio (SINR),
\begin{align}
 {\mathbf{\xi}_{c,k}}(\mathbf{w}_c,\mathbf{w}_k,\mathbf{h}_k ) = 
\frac{  | \mathbf{h}_{k}^\dag \mathbf{w}_{c}|^2}
{ \sum_{j=1}^{K}| \mathbf{h}_{k}^\dag \mathbf{w}_{j} |^2
+ \sigma_{k}^{2} } ,
 \ \forall k\in \mathcal{K}.
\end{align}   
For the sake of simplicity, we let $D_{c,k}=\frac{Q^{-1}(\epsilon_k )}{\sqrt{L_{c,k}}  }$, $L_{c,k}=L$.

Once $s_c$ is successfully decoded and removed from $y_k$, the desired private stream $s_k$ can be decoded by treating the other private streams $s_j$, $j \in \mathcal{K}$, $j \neq k$ as noise. Similarly, the achievable  rate of decoding  $s_k$ at user $k$ can be written as
\begin{align}
R_{p,k}(\mathbf{w}_k,\mathbf{h}_k )=&  \ \mathrm{ln} (1+ { \mathbf{\xi}_{p,k} } )  -
 \sqrt{\Lambda( \mathbf{\xi}_{p,k} ) }\ \frac{ Q^{-1}(\epsilon_k )}{\sqrt{L_{p,k}} } ,\notag \\
& \quad  \quad \quad  \quad  \quad  \quad \quad  \quad  \quad  \quad \quad  \quad  \forall k\in \mathcal{K},\
\end{align} 
where
\begin{align}
 {\mathbf{\xi}_{p,k}}(\mathbf{w}_k,\mathbf{h}_k ) = 
\frac{  | \mathbf{h}_{k}^\dag \mathbf{w}_{k}|^2}
{ \sum_{j=1,j\neq k}^{K}| \mathbf{h}_{k}^\dag \mathbf{w}_{j} |^2
+ \sigma_{k}^{2} } ,
 \ \forall k\in \mathcal{K},
\end{align} 
 $L_{p,k}$ is the  blocklength of stream $s_k$. For the sake of simplicity, we let  $D_{p,k}=\frac{Q^{-1}(\epsilon_k )}{\sqrt{L_{p,k}}  }$, $L_{p,k}=L$.

Since the common stream $s_c$ is broadcast to and decoded by all the receivers, the  achievable data rate $R_c$ of the common stream $s_c$ is actually limited by the minimum common rate among users, i.e.,
\begin{align}\label{minimu_c}
 R_c \leq \min_{k\in \mathcal{K}} R_{c,k}(\mathbf{w}_c,\mathbf{w}_k,\mathbf{h}_k ) .
\end{align} 
Note that only the portion of common rate, $\frac{|W_{c,k}|}{\sum_{j \in \mathcal{K}} |W_{c,j}|} R_c$  is intended for user $k$, where $|W|$ denotes the length of message $W$. For easy denotation, we define
a minimum common rate among users, that is,
\begin{align}
 c_k =\frac{|W_{c,k}|}{\sum_{j \in \mathcal{K}} |W_{c,j}|} R_c \notag
\end{align} 
as the common data rate or rate-splitting factor of user $k$, which satisfies $\sum_{ j \in \mathcal{K}} c_j = R_c$ and $c_j \geq 0$, $j \in \mathcal{K}$. Then \eqref{minimu_c} can be rewritten as 
\begin{align}\label{common_rate}
 \sum_{ j \in \mathcal{K}} c_j  \leq \min_{k\in \mathcal{K}} R_{c,k}(\mathbf{w}_c,\mathbf{w}_k,\mathbf{h}_k ) .
\end{align}

In  practical  multi-user MISO  networks,  due to the channel estimation error   or the quantization feedback error, the acquired imperfect CSIT can dramatically degrade the  xURLLC system performance. To guarantee the   transmission latency and  reliability, we propose a robust  beamforming  and rate-splitting design in FBL regime under imperfect CSIT. The well-known norm-bounded error model is adopted \cite{Robust2016liao}, in which the channel estimation error  $\Delta \mathbf{h}_{k}$ is bounded by $||\Delta \mathbf{h}_{k}|| \leq \delta_k$, where $\delta_k$ indicates the accuracy of estimation. Denoting the estimated channel coefficient as $\hat{\mathbf{h}}_{k}$, the actual channel coefficient $\mathbf{h}_{k}$ is within a ball, i.e., \begin{align}
\mathbf{h}_{k} \in \mathcal{H}_{k}=\left\{ \hat{\mathbf{h}}_{k}+\Delta \mathbf{h}_{k}|\left\| \Delta \mathbf{h}_{k} \right\| \leq \delta_k\right\},\ \forall k \in \mathcal{K}.
\end{align} 
To incorporate the communication scenarios where CSIT errors decay with increased signal-to-noise ratio (SNR) through increasing the number of feedback bits, the channel estimation error  power $\delta_{k}^2$ is allowed to scale as $O($SNR$^{-\alpha_k}$) where $\alpha_k \triangleq \lim_{P_{\max}\to \infty} - \frac{\mathrm{log}(\delta_{k}^2)}{\mathrm{log}(P_{\max})}$ denotes the CSIT quality factor as SNR grows large \
\cite{Joudeh2016sumrate,joudeh2016robust}.  $\alpha_k  \to  \infty$ represents perfect CSIT (i.e., $\delta_{k} =0$) resulting from an infinitely high number of feedback bits,  while  $\alpha_k  =0$ represents fixed feedback bits and fixed CSIT quality. In  the DoF sense, the scaling exponents are truncated such that $\alpha_k \in [0, 1]$. 

Based on this model, the robust design should ensure that certain performance can be achieved over all possible channels in the uncertain region. Specifically, due to the uncertainty of CSIT, the  achievable data rate falls into a bounded uncertainty region. The worst achievable data rates for the common  and private streams of user $k$ are expressed as
\begin{align} 
\bar{R}_{c,k}(\mathbf{w}_c,\mathbf{w}_k ) \triangleq  \min_{ \mathbf{h}_k \in \mathcal{H}_k} R_{c,k}(\mathbf{w}_c,\mathbf{w}_k,\mathbf{h}_k ), \ k \in \mathcal{K},
\end{align}
\begin{align} 
\bar{R}_{p,k}(\mathbf{w}_k ) \triangleq  \min_{ \mathbf{h}_k \in \mathcal{H}_k} R_{p,k}(\mathbf{w}_k,\mathbf{h}_k ), \ k \in \mathcal{K},
\end{align}
respectively. In this case, the  achievable data rate $R_c$ or $\sum_{ j \in \mathcal{K}} c_j $ of the common stream $s_c$ should no bigger than the minimum common rate for an arbitrary user $k \in \mathcal{K}$  with any channel uncertainty $\{ \Delta \mathbf{h}_{k} \}$, i.e.,
\begin{align}
 \sum_{ j \in \mathcal{K}} c_j  \leq   \min_{ \mathbf{h}_k \in \mathcal{H}_k} R_{c,k}(\mathbf{w}_c,\mathbf{w}_k,\mathbf{h}_k ), \ k \in \mathcal{K},
\end{align} 
or in another  form,
\begin{align}\label{c_imperfect}
 \sum_{ j \in \mathcal{K}} c_j  \leq   R_{c,k}(\mathbf{w}_c,\mathbf{w}_k,\mathbf{h}_k ) ,\ \forall \mathbf{h}_k \in \mathcal{H}_k, \ \forall k \in \mathcal{K}.
\end{align}

In summary, the total achievable data rate of user $k$ includes the  common rate $c_k$ and the private rate $R_{p,k}(\mathbf{w}_k,\mathbf{h}_k )$,  i.e., $R_k = c_k + R_{p,k}(\mathbf{w}_k,\mathbf{h}_k ) $, $k \in \mathcal{K}$. 

\section{Problem Formulation and Problem Transformation} \label{Formulation}
Based on the above mentioned  transmission scheme in Section II,  the minimum user rate maximization problem is formulated in this section   under the premise of ensuring the requirements of transmission latency and transmission reliability. To solve the formulated  non-convex problem, we transform it into an equivalent one which is more tractable by properly introducing auxiliary variables.

\subsection{Max-Min Rate Optimization Problem Formulation}
Based on the system model and the  transmission scheme in Section II, our objective is to maximize the  minimum user rate  over all possible channels in the uncertain region under the premise of ensuring the requirements of  the packet transmission latency  $L_k/\mathrm{B}$ (i.e., the duration of the continuous-time signal) \cite{fu2021resource} and  the transmission reliability  $\epsilon_k$ of all the xURLLC users, constrained by the power budget. To this end, the common beamforming vector $\mathbf{w}_{c}$, the private beamforming vectors $\{ \mathbf{w}_{k}\}$, $ k\in \mathcal{K} $, and the rate-splitting vector $\mathbf{c} = [c_1,...,c_k,...,c_K]^T$ are jointly optimized. For  given blocklength $L_k$ and   transmission reliability  $\epsilon_k$, $ k \in \mathcal{K}$, i.e., given the user requirements,  the robust beamforming and rate-splitting optimization problem in the FBL regime can be formulated as
\begin{subequations} \label{eq_p_problem}
\begin{align} 
& \max_{\{ \mathbf{w}_{k} \},\mathbf{w}_{c},\mathbf{c} } \   \min_{k \in \mathcal{K}} \ [ c_k + \min_{\mathbf{h}_k \in \mathcal{H}_k}R_{p,k}(\mathbf{w}_k,\mathbf{h}_k ) ]
 \\
  \text{s.t.} 
& \    \sum_{ j \in \mathcal{K}} c_j  \leq   R_{c,k}(\mathbf{w}_c,\mathbf{w}_k,\mathbf{h}_k ) ,\ \forall \mathbf{h}_k \in \mathcal{H}_k, \ \forall k \in \mathcal{K}, \label{pp_1} \\
& \ \mathbf{w}_{c}^\dag \mathbf{w}_{c}+\sum_{k=1}^{K} \mathbf{w}_{k}^\dag \mathbf{w}_{k} \leq P_{\max}, \label{pp_2}\\
& \    c_k \geq 0, \ \forall k\in \mathcal{K}.\label{pp_3} 
\end{align}
\end{subequations}
 Constraint  \eqref{pp_1}  ensures that the common stream $s_c$ can be successfully decoded by all the xURLLC users. Constraint  \eqref{pp_2} is the total transmission power constraint. We have an observation from \eqref{eq_p_problem} that the variables $\{ \mathbf{w}_{k} ,\mathbf{w}_{c},\mathbf{c}\} $ are coupled with each other in the objective function and  constraint \eqref{pp_1}. In addition,  the asymptotic capacity based on the finite blocklength coding theory is a complex and non-convex function of the beamforming vectors. Hence, the optimization problem  \eqref{eq_p_problem} is non-convex due to the non-convexity of the constraint set.   As a result,  it is  computationally intractable to find the global  optimal solution for  \eqref{eq_p_problem} due to its non-convex constraint set and the infinite constraints. 
  
  By introducing auxiliary variable $t$, the optimization problem \eqref{eq_p_problem}  can be equivalently reformulated into the following one
 \begin{subequations} \label{eq_primal_problem}
\begin{align} 
& \max_{\{ \mathbf{w}_{k} \},\mathbf{w}_{c},\mathbf{c},t } \   t \\
  \text{s.t.} 
  &  \ \  t \leq c_k + R_{p,k}(\mathbf{w}_k,\mathbf{h}_k ) ,\ \forall \mathbf{h}_k \in \mathcal{H}_k, \ \forall k\in \mathcal{K},\label{p_5} \\
& \    \sum_{ j \in \mathcal{K}} c_j  \leq   R_{c,k}(\mathbf{w}_c,\mathbf{w}_k,\mathbf{h}_k ) ,\ \forall \mathbf{h}_k \in \mathcal{H}_k, \ \forall k \in \mathcal{K}, \label{p_1} \\
& \ \mathbf{w}_{c}^\dag \mathbf{w}_{c}+\sum_{k=1}^{K} \mathbf{w}_{k}^\dag \mathbf{w}_{k} \leq P_{\max}, \label{p_2}\\
& \    c_k \geq 0, \ \forall k\in \mathcal{K}.\label{p_3} 
\end{align}
\end{subequations} 
  The transformed problem is still non-convex due to the
non-convex constraints \eqref{p_5}, \eqref{p_1}.

\subsection{Problem Transformation}
The complexity of finding the global minimum of \eqref{eq_primal_problem} is prohibitively high. To make the non-convex
 problem more tractable, in the following we transform \eqref{eq_primal_problem} 
 into an equivalent one by introducing some auxiliary variables.
 
 By introducing vector $\boldsymbol{\beta}_c = [\beta_{c,1},...,\beta_{c,k},...,\beta_{c,K}]^T$ with its element $\beta_{c,k}$ representing the SINR of common stream at user $k$, constraint \eqref{p_1} is equivalent to 
 \begin{subnumcases}{\eqref{p_1}\Leftrightarrow}
 - \ \mathrm{ln} (1+ \beta_{c,k} )  +\sum_{ j \in \mathcal{K}} c_j  \leq  -D_{c,k} \notag \\
 \quad \ \ \quad \quad   \sqrt{1-(1+\beta_{c,k})^{-2}}   , \ \forall k\in \mathcal{K}, \label{eq_reform_1_1}  \\
\beta_{c,k} \leq \frac{  | \mathbf{h}_{k}^\dag \mathbf{w}_{c}|^2}
{ \sum_{j=1}^{K}| \mathbf{h}_{k}^\dag \mathbf{w}_{j} |^2
+ \sigma_{k}^{2} } ,\ \forall \mathbf{h}_k \in \mathcal{H}_k, \notag \\
 \quad \quad \quad \quad \quad \quad \quad \quad \quad \quad \quad \quad   \forall k\in \mathcal{K}.\label{eq_reform_1_2}
\end{subnumcases}
It can be found that constraint \eqref{eq_reform_1_1} is in the form of difference of convex (DC) programming, and \eqref{eq_reform_1_2} contains infinite constraints due to the norm-bounded error model. To tackle  \eqref{eq_reform_1_2}, we further introduce two auxiliary variable vectors, $\mathbf{x}_c = [x_{c,1},...,x_{c,k},...,x_{c,K}]^T$ and $\mathbf{y}_c = [y_{c,1},...,y_{c,k},...,y_{c,K}]^T$, \eqref{eq_reform_1_2}  is equivalent to
 \begin{subnumcases}{\eqref{eq_reform_1_2}\Leftrightarrow}
 e^{x_{c,k}} \leq | \mathbf{h}_{k}^\dag \mathbf{w}_{c}|^2, \ \forall \mathbf{h}_k \in \mathcal{H}_k,\ \forall k\in \mathcal{K},\ \  \ \label{eq_reform_1_2_1}  \\
   \sum_{j=1}^{K}| \mathbf{h}_{k}^\dag \mathbf{w}_{j} |^2
+ \sigma_{k}^{2} \leq   e^{y_{c,k}}  , \ \forall \mathbf{h}_k \in \mathcal{H}_k,\notag \\
\quad  \quad \quad \quad \quad   \quad \quad \quad  \quad \quad \quad \ \forall k\in \mathcal{K}, \label{eq_reform_1_2_2}  \\
\beta_{c,k} \leq   e^{x_{c,k}-y_{c,k}} ,\ \forall k\in \mathcal{K}.\label{eq_reform_1_2_3}
\end{subnumcases}
\eqref{eq_reform_1_2_1} and \eqref{eq_reform_1_2_2} are still infinite constraints
and not easy to deal with. We introduce another two auxiliary variable vectors, $\mathbf{t}_c = [t_{c,1},...,t_{c,k},...,t_{c,K}]^T$ and $\mathbf{q}_c = [q_{c,1},...,q_{c,k},...,q_{c,K}]^T$, \eqref{eq_reform_1_2_1} and \eqref{eq_reform_1_2_2} are equivalent to
 \begin{subnumcases}{\eqref{eq_reform_1_2_1}\eqref{eq_reform_1_2_2}\Leftrightarrow}
 e^{x_{c,k}} \leq t_{c,k} , \  \forall k\in \mathcal{K}, \label{eq_reform_1_2_1_1}  \\
 t_{c,k} \leq | \mathbf{h}_{k}^\dag \mathbf{w}_{c}|^2, \ \forall \mathbf{h}_k \in \mathcal{H}_k,\ \forall k\in \mathcal{K}, \quad \quad \ \ \label{eq_reform_1_2_1_2} \\
   q_{c,k}  \leq e^{y_{c,k}}, \ \forall k\in \mathcal{K}, \ \label{eq_reform_1_2_1_3}  \\
 \sum_{j=1}^{K}| \mathbf{h}_{k}^\dag \mathbf{w}_{j} |^2
+ \sigma_{k}^{2} \leq   q_{c,k} ,\ \forall \mathbf{h}_k \in \mathcal{H}_k,\notag \\
\quad \quad \quad \quad \quad \quad \quad \quad  \quad \quad  \ \forall k\in \mathcal{K},\label{eq_reform_1_2_1_4}
\end{subnumcases}
where \eqref{eq_reform_1_2_1_1} is affine function, \eqref{eq_reform_1_2_1_3} is in the form of DC. For the infinite constraints \eqref{eq_reform_1_2_1_2} and \eqref{eq_reform_1_2_1_4}, we adopt S-Procedure method to convert them into a linear matrix inequality (LMI), whose number of constraints is finite \cite{boyd2004convex}. To be  specific, let $\mathbf{W}_k=\mathbf{w}_k \mathbf{w}_k^\dag \succeq \mathbf{0}$, $\mathrm{rank}(\mathbf{W}_k)= 1$,  $\mathbf{W}_c=\mathbf{w}_c \mathbf{w}_c^\dag \succeq \mathbf{0}$, $\mathrm{rank}(\mathbf{W}_c)= 1$,  \eqref{eq_reform_1_2_1_2} and \eqref{eq_reform_1_2_1_4} can be converted into the following LMI constraints,
\begin{equation}\label{eq_PSD_1}
\boldsymbol{\Gamma}_k(\mathbf{W}_c)\succeq\mathbf{0},\ k\in\mathcal{K},
\end{equation}
\begin{equation}\label{eq_PSD_2}
\boldsymbol{\Psi}_k(\mathbf{W}_k)\succeq\mathbf{0},\ k\in\mathcal{K},
\end{equation}
respectively, where $\boldsymbol{\Gamma}_k(\mathbf{W}_c)$ and $\boldsymbol{\Psi}_k(\mathbf{W}_k)$ are positive semidefinite (PSD) yet hermitian matrices that can be expressed as
\begin{align}
&\boldsymbol{\Gamma}_k(\mathbf{W}_c)=
\left[ 
\begin{array}{ccc}
\lambda_{c,k} \mathbf{I}_M +\mathbf{W}_c & \mathbf{W}_c \hat{\mathbf{h}}_{k} \\
\hat{\mathbf{h}}_{k}^\dag \mathbf{W}_c ^\dag & \hat{\mathbf{h}}_{k}^\dag \mathbf{W}_c \hat{\mathbf{h}}_{k}-\lambda_{c,k} \delta_k^2- t_{c,k}  \\
\end{array} 
\right],\nonumber
\end{align}

\begin{align}
&\boldsymbol{\Psi}_k(\mathbf{W}_k)=
\left[ 
\begin{array}{ccc}
\bar{\lambda}_{c,k} \mathbf{I}_M -\mathbf{Z}_k  & -\mathbf{Z}_k \hat{\mathbf{h}}_{k} \\
-\hat{\mathbf{h}}_{k}^\dag \mathbf{Z}_k ^\dag & -\hat{\mathbf{h}}_{k}^\dag \mathbf{Z}_k \hat{\mathbf{h}}_{k}-\bar{\lambda}_{c,k} \delta_k^2+ q_{c,k}  \\
\end{array} 
\right],\nonumber
\end{align}
respectively, where $\mathbf{Z}_k = \sum_{j=1}^{K}\mathbf{W}_k$.

Focusing  on constraint \eqref{eq_reform_1_2_3}, the function $ f_1(x_{c,k},y_{c,k})=e^{x_{c,k}-y_{c,k}}$ is twice differentiable, and all its second partial derivatives exist and are continuous over the domain of $f_1$. Then the Hessian $\bigtriangledown^2 f_1(x_{c,k},y_{c,k})$ of $f_1$ is expressed as
\begin{align}
\bigtriangledown^2 f_1(x_{c,k},y_{c,k})&=
\left[ 
\begin{array}{ccc}
\frac{\partial^2 f_1 }{\partial^2 x_{c,k} } & \frac{\partial^2 f_1 }{\partial x_{c,k} \ \partial y_{c,k}}  \\
\frac{\partial^2 f_1 }{\partial y_{c,k} \ \partial x_{c,k}}  & \frac{\partial^2 f_1 }{\partial^2 y_{c,k}}  \\
\end{array} 
\right]\nonumber \\
&=
\left[ 
\begin{array}{ccc}
e^{x_{c,k}-y_{c,k}} & -e^{x_{c,k}-y_{c,k}}  \\
-e^{x_{c,k}-y_{c,k}}   & e^{x_{c,k}-y_{c,k}}   \\
\end{array} 
\right].\nonumber 
\end{align}
It is obvious that the matrix is semidefinite. Then we get that the function $ f_1(x_{c,k},y_{c,k})=e^{x_{c,k}-y_{c,k}}$ is convex. That is to say, constraint \eqref{eq_reform_1_2_3} is in the form of DC.

So far the constraint \eqref{p_1} in the original problem \eqref{eq_primal_problem}  has been transformed into the constraints of DC form \eqref{eq_reform_1_1}, \eqref{eq_reform_1_2_3} and \eqref{eq_reform_1_2_1_3}, the  affine constraints \eqref{eq_reform_1_2_1_1}, as well as the LMI constraints \eqref{eq_PSD_1} and \eqref{eq_PSD_2}.

Similarly, the constraint \eqref{p_5} in the original problem \eqref{eq_primal_problem} can also be transformed into the following constraints equivalently through the same approach,
\begin{subnumcases}{\eqref{p_2}\Leftrightarrow}
 - \ \mathrm{ln} (1+ \beta_{p,k} ) +\ t - c_k \leq   -D_{p,k} \notag \\
 \quad \quad \quad      \sqrt{1-(1+\beta_{p,k})^{-2}}  , \ \forall k\in \mathcal{K}, \label{2_eq_reform_1_2_1_0} \\
 e^{x_{p,k}} \leq t_{p,k} , \  \forall k\in \mathcal{K}, \label{2_eq_reform_1_2_1_1}  \\
    q_{p,k} \leq e^{y_{p,k}}, \ \forall k\in \mathcal{K}, \label{2_eq_reform_1_2_1_2}  \\
\boldsymbol{\Omega}_k(\mathbf{W}_k)\succeq\mathbf{0},\ k\in\mathcal{K},\label{2_eq_reform_1_2_1_3} \\
\boldsymbol{\Theta}_k(\mathbf{W}_k)\succeq\mathbf{0},\ k\in\mathcal{K},\label{2_eq_reform_1_2_1_4}  \\
\beta_{p,k} \leq   e^{x_{p,k}-y_{p,k}} ,\ \forall k\in \mathcal{K},\label{2_eq_reform_1_2_1_5} 
\end{subnumcases}
where $\boldsymbol{\beta}_p = [\beta_{p,1},...,\beta_{p,k},...,\beta_{p,K}]^T$, $\mathbf{x}_p = [x_{p,1},...,x_{p,k},...,x_{p,K}]^T$, $\mathbf{y}_p = [y_{p,1},...,y_{p,k},...,y_{p,K}]^T$, $\mathbf{t}_p = [t_{p,1},...,t_{p,k},...,t_{p,K}]^T$ and $\mathbf{q}_p = [q_{p,1},...,q_{p,k},...,q_{p,K}]^T$ are the introduced auxiliary variable vectors related to the private streams. $\boldsymbol{\Omega}_k(\mathbf{W}_k)$ and $\boldsymbol{\Theta}_k(\mathbf{W}_k)$ are expressed  as

\begin{align}
&\boldsymbol{\Omega}_k(\mathbf{W}_k)=
\left[ 
\begin{array}{ccc}
\lambda_{p,k} \mathbf{I}_M +\mathbf{W}_k & \mathbf{W}_k \hat{\mathbf{h}}_{k} \\
\hat{\mathbf{h}}_{k}^\dag \mathbf{W}_k ^\dag & \hat{\mathbf{h}}_{k}^\dag \mathbf{W}_k \hat{\mathbf{h}}_{k}-\lambda_{k,k} \delta_k^2- t_{k,k}  \\
\end{array} 
\right],\nonumber
\end{align}

\begin{align}
&\boldsymbol{\Theta}_k(\mathbf{W}_k)=
\left[ 
\begin{array}{ccc}
\bar{\lambda}_{p,k} \mathbf{I}_M -\mathbf{N}_k  & -\mathbf{N}_k \hat{\mathbf{h}}_{k} \\
-\hat{\mathbf{h}}_{k}^\dag \mathbf{N}_k ^\dag & -\hat{\mathbf{h}}_{k}^\dag \mathbf{N}_k \hat{\mathbf{h}}_{k}-\bar{\lambda}_{p,k} \delta_k^2+ q_{p,k}  \\
\end{array} 
\right],\nonumber
\end{align}
respectively, where $\mathbf{N}_k = \sum_{j=1,j\neq k}^{K}\mathbf{W}_k$.

For simplicity, let $\mathbf{z}$ = $[ \mathbf{W}_{k} ,\mathbf{W}_{c},\mathbf{c},t,\boldsymbol{\beta}_c ,\mathbf{x}_{c} ,\mathbf{y}_{c},\mathbf{t}_c,\mathbf{q}_c,$ $\boldsymbol{\beta}_p ,\mathbf{x}_{p} ,\mathbf{y}_{p} ,\mathbf{t}_p,\mathbf{q}_p]^T$. 
Therefore, problem \eqref{eq_primal_problem} is equivalently transformed into
\begin{align} \label{eq_primal_problem_2}
&  \max_{\mathbf{z}} \  
 t  \\
  \text{s.t.} 
& \ \  \ \eqref{eq_reform_1_1} , \eqref{eq_reform_1_2_3}, \eqref{eq_reform_1_2_1_1}, \eqref{eq_reform_1_2_1_3},  \eqref{eq_PSD_1}, \eqref{eq_PSD_2},  \notag \\
& \ \ \  \eqref{2_eq_reform_1_2_1_0}-\eqref{2_eq_reform_1_2_1_5},\eqref{p_2},\eqref{p_3}, \notag \\ 
& \ \ \ \mathrm{rank}(\mathbf{W}_c)= 1,\mathrm{rank}(\mathbf{W}_k)= 1,\ \forall k\in \mathcal{K}\notag.
\end{align}

As the constraints \eqref{eq_reform_1_1}, \eqref{eq_reform_1_2_3},     \eqref{eq_reform_1_2_1_3}, \eqref{2_eq_reform_1_2_1_0} ,\eqref{2_eq_reform_1_2_1_2}, and \eqref{2_eq_reform_1_2_1_5} are in  the form of DC, the  equivalently transformed problem \eqref{eq_primal_problem_2} is a rank-one constrained DC programming one which is still non-convex.

\section{ Robust Beamforming and Rate-Splitting Design } \label{Design}
In this section, the transformed rank-one constrained DC programming \eqref{eq_primal_problem_2} is 
addressed under the imperfect instantaneous CSIT case. To solve it, the CCCP is utilized to achieve a local minimum of  \eqref{eq_primal_problem_2}. Furthermore,  a feasible point search algorithm is proposed to initialize the solving process.

\subsection{ Proposed Iterative Algorithm } \label{imperfect CSI}
In \eqref{eq_primal_problem_2}, the approach to transform DC non-convex constraints \eqref{eq_reform_1_1}, \eqref{eq_reform_1_2_3},     \eqref{eq_reform_1_2_1_3}, \eqref{2_eq_reform_1_2_1_0} ,\eqref{2_eq_reform_1_2_1_2} and \eqref{2_eq_reform_1_2_1_5} into convex ones is to approximate the convex terms on the right hand into concave  ones. To achieve that, we adopt the CCCP-based method to iteratively approximate the original non-convex feasible set defined in these constraints by its convex subset around certain feasible point \cite{sriperumbudur2009convergence,smola2005kernel, cheng2012joint}. To be specific, we try to approximate the convex functions $f_2(\beta_{c,k}) =-D_{c,k} \sqrt{1-(1+\beta_{c,k})^{-2}} $ in \eqref{eq_reform_1_1}, $f_1(x_{c,k},y_{c,k})=e^{x_{c,k}-y_{c,k}} $ in \eqref{eq_reform_1_2_3}, $f_3(y_{c,k})=e^{y_{c,k}} $ in \eqref{eq_reform_1_2_1_3}, $f_4(\beta_{p,k}) =-D_{p,k} \sqrt{1-(1+\beta_{p,k})^{-2}}$ in \eqref{2_eq_reform_1_2_1_0}, $f_5(y_{p,k})=e^{y_{p,k}} $ in \eqref{2_eq_reform_1_2_1_2}, and $f_6(x_{p,k},y_{p,k})=e^{x_{p,k}-y_{p,k}} $  in \eqref{2_eq_reform_1_2_1_5} by their first order Taylor expansions, which are apparently affine functions (both convex and concave). Then these non-convex constraints are transferred into convex ones. Through iteratively solving the resulting convex approximation, a local minimum is expected to be finally achieved.

Recall that the first-order Taylor series expansion of a function $f:\mathbb{R}^n  \to \mathbb{R}$ at point $\mathbf{x}^{(i)}$  is given by 
\begin{align}
&\hat{ f}(\mathbf{x}^{(i)},\mathbf{x})=f(\mathbf{x}^{(i)})+\bigtriangledown f(\mathbf{x}^{(i)})^T (\mathbf{x}-\mathbf{x}^{(i)}),
\end{align}
where $\bigtriangledown f (\mathbf{x}^{(i)})$ denotes the gradient of $f(\mathbf{x})$ with respect to vector $\mathbf{x}$ evaluated at the point $\mathbf{x}^{(i)}$. For convex functions, $f (\mathbf{x}^{(i)}, \mathbf{x}) \geq \hat{f} (\mathbf{x}^{(i)}, \mathbf{x})$.

So we approximate the convex function $f_2(\beta_{c,k})=-D_{c,k} \sqrt{1-(1+\beta_{c,k})^{-2}}$  using the
first-order lower approximation $\hat{f_2} (\beta_{c,k})$ at point $\beta_{c,k}^{(i)}$ in the $i$-th iteration \cite{chi2017convex}, i.e.,
\begin{align}\label{f_2}
\hat{f_2} (\beta_{c,k}^{(i)},\beta_{c,k})=
&-D_{c,k} \sqrt{1-(1+\beta_{c,k}^{(i)})^{-2}} \notag\\ 
&-\frac{D_{c,k}(\beta_{c,k}-\beta_{c,k}^{(i)}) }{\sqrt{1-(1+\beta_{c,k}^{(i)})^{-2}}\ (1+\beta_{c,k}^{(i)})^{3}},
\end{align}

Similarly, $f_4(\beta_{p,k}) =-D_{p,k} \sqrt{1-(1+\beta_{p,k})^{-2}}$ in \eqref{2_eq_reform_1_2_1_0} is approximated as 
\begin{align}\label{f_4}
\hat{f_4} (\beta_{p,k}^{(i)},\beta_{p,k})=
&-D_{p,k} \sqrt{1-(1+\beta_{p,k}^{(i)})^{-2}} \notag\\ 
&-\frac{D_{p,k}(\beta_{p,k}-\beta_{p,k}^{(i)}) }{\sqrt{1-(1+\beta_{p,k}^{(i)})^{-2}}\ (1+\beta_{p,k}^{(i)})^{3}}.
\end{align}
Obviously,  the first-order lower approximations $\hat{f_2}$ and  $\hat{f_4}$ are affine functions. For the convex function $f_1(x_{c,k},y_{c,k})=e^{x_{c,k}-y_{c,k}} $ and $f_6(x_{p,k},y_{p,k})=e^{x_{p,k}-y_{p,k}} $, they can  be approximated as 
\begin{align}\label{f_1}
&\hat{f_1} ([x_{c,k}^{(i)}\ y_{c,k}^{(i)}]^T,[x_{c,k} \ y_{c,k}]^T) \notag\\
=& \ e^{x_{c,k}^{(i)}-y_{c,k}^{(i)}} + 
\left[ 
\begin{array}{ccc}
e^{x_{c,k}^{(i)}-y_{c,k}^{(i)}} & -e^{x_{c,k}^{(i)}-y_{c,k}^{(i)}}  \\
\end{array} 
\right]\nonumber 
\left[ 
\begin{array}{ccc}
x_{c,k}-x_{c,k}^{(i)}  \\
y_{c,k}-y_{c,k}^{(i)} \\
\end{array} 
\right] \notag\\
=& \ e^{x_{c,k}^{(i)}-y_{c,k}^{(i)}} \  [  1+x_{c,k}-x_{c,k}^{(i)}-y_{c,k}+y_{c,k}^{(i)} ],
\end{align}
\begin{align}\label{f_6}
&\hat{f_6} ([x_{p,k}^{(i)}\ y_{p,k}^{(i)}]^T,[x_{p,k} \ y_{p,k}]^T) \notag\\
=& \ e^{x_{p,k}^{(i)}-y_{p,k}^{(i)}} + 
\left[ 
\begin{array}{ccc}
e^{x_{p,k}^{(i)}-y_{p,k}^{(i)}} & -e^{x_{p,k}^{(i)}-y_{p,k}^{(i)}}  \\
\end{array} 
\right]\nonumber 
\left[ 
\begin{array}{ccc}
x_{p,k}-x_{p,k}^{(i)}  \\
y_{p,k}-y_{p,k}^{(i)} \\
\end{array} 
\right]\notag \\
=&\  e^{x_{p,k}^{(i)}-y_{p,k}^{(i)}} \  [  1+x_{p,k}-x_{p,k}^{(i)}-y_{p,k}+y_{p,k}^{(i)} ],
\end{align}
which are also affine functions.

At last, the convex function $f_3(y_{c,k})=e^{y_{c,k}} $ in \eqref{eq_reform_1_2_1_3} and  $f_5(y_{p,k})=e^{y_{p,k}} $ in \eqref{2_eq_reform_1_2_1_2} can be approximated as the following affine functions,
\begin{align}\label{f_3}
&\hat{f_3} ( y_{c,k}^{(i)},y_{c,k}) 
= e^{y_{c,k}^{(i)}} (1+y_{c,k}-y_{c,k}^{(i)}),
\end{align}
\begin{align}\label{f_5}
&\hat{f_5} ( y_{p,k}^{(i)},y_{p,k}) 
= e^{y_{p,k}^{(i)}} (1+y_{p,k}-y_{p,k}^{(i)}),
\end{align}
respectively.

Consequently, the DC non-convex constraints \eqref{eq_reform_1_1}, \eqref{eq_reform_1_2_3},     \eqref{eq_reform_1_2_1_3}, \eqref{2_eq_reform_1_2_1_0} ,\eqref{2_eq_reform_1_2_1_2} and \eqref{2_eq_reform_1_2_1_5} in problem \eqref{eq_primal_problem_2} are approximated as the following convex inequalities  at point $\mathbf{x}^{(i)}=\{ \beta_{c,k}^{(i)},x_{c,k}^{(i)}, y_{c,k}^{(i)}, \beta_{p,k}^{(i)},x_{p,k}^{(i)}, y_{p,k}^{(i)}\}^T$ in the $i$-th iteration:
 \begin{subnumcases}
 {}
- \ \mathrm{ln} (1+ \beta_{c,k} )  +\sum_{ j \in \mathcal{K}} c_j     \leq  
 \hat{f_2} (\beta_{c,k}^{(i)},\beta_{c,k}), \ \forall k\in \mathcal{K},\label{convex_1}\\
 \beta_{c,k} \leq  \hat{f_1} ([x_{c,k}^{(i)}\ y_{c,k}^{(i)}]^T,[x_{c,k} \ y_{c,k}]^T) ,\ \forall k\in \mathcal{K},\label{convex_2}\\
  q_{c,k} \leq \hat{f_3} ( y_{c,k}^{(i)},y_{c,k})  , \ \forall k\in \mathcal{K},\label{convex_3}\\ 
 - \ \mathrm{ln} (1+ \beta_{p,k} ) +t - c_k \leq
\hat{f_4} (\beta_{p,k}^{(i)},\beta_{p,k})  ,  \forall k\in \mathcal{K}, \ \ \ \ \ \ \ \label{convex_4}\\ q_{p,k} \leq \hat{f_5} ( y_{p,k}^{(i)},y_{p,k})  , \ \forall k\in \mathcal{K},\label{convex_5}\\
 \beta_{p,k} \leq   \hat{f_6} ([x_{p,k}^{(i)}\ y_{p,k}^{(i)}]^T,[x_{p,k} \ y_{p,k}]^T),\ \forall k\in \mathcal{K}.\label{convex_6}
\end{subnumcases}

As $   {f_1}  \geq   \hat{f_1}  $, ${f_2}   \geq  \hat{f_2} $,  ${f_3} \geq \hat{f_3} $,  ${f_4} \geq \hat{f_4} $, ${f_5}  \geq \hat{f_5} $,  ${f_6}  \geq \hat{f_6} $ \cite{boyd2004convex}, the feasible convex set defined in \eqref{convex_1}-\eqref{convex_6} is a subset of that in \eqref{eq_reform_1_1}, \eqref{eq_reform_1_2_3},     \eqref{eq_reform_1_2_1_3}, \eqref{2_eq_reform_1_2_1_0},\eqref{2_eq_reform_1_2_1_2} and \eqref{2_eq_reform_1_2_1_5}. Thus the latter can be approximated by the former around the current feasible point $\{ \beta_{c,k}^{(i)},x_{c,k}^{(i)}, y_{c,k}^{(i)}, \beta_{p,k}^{(i)},x_{p,k}^{(i)}, y_{p,k}^{(i)}\}^T$. 
Based on the preliminaries above, the  DC programming \eqref{eq_primal_problem_2} can be approximated as a rank-one constrained  optimization problem in the $i$-th iteration as follows 
\begin{align} \label{eq_rank_1_convex_problem} 
& \ \ \ \max_{ \mathbf{z} } \  
 t \\
  \text{s.t.} 
& \ \   \eqref{eq_reform_1_2_1_1},  \eqref{eq_PSD_1}, \eqref{eq_PSD_2},    \eqref{2_eq_reform_1_2_1_1}, \eqref{2_eq_reform_1_2_1_3},\eqref{2_eq_reform_1_2_1_4},\eqref{p_2},\eqref{p_3}, \notag \\
& \ \ \  \eqref{convex_1}-\eqref{convex_6}\notag.\\
& \ \ \ \mathrm{rank}(\mathbf{W}_c)= 1,\mathrm{rank}(\mathbf{W}_k)= 1,\  \forall k\in \mathcal{K}\notag.
\end{align} 
\eqref{eq_rank_1_convex_problem} is still non-convex due to the rank-one constraints. We can adopt the semi-positive definite relaxation (SDR) and  Gaussian randomization method \cite{Robust2016liao} to generate a feasible solution to \eqref{eq_rank_1_convex_problem}. We first drop the rank-one constraints and  obtain the following convex problem, 
\begin{align} \label{eq_convex_problem} 
& \ \ \  \max_{ \mathbf{z}} \  
 t \\
  \text{s.t.} 
& \ \   \eqref{eq_reform_1_2_1_1},  \eqref{eq_PSD_1}, \eqref{eq_PSD_2},    \eqref{2_eq_reform_1_2_1_1}, \eqref{2_eq_reform_1_2_1_3},\eqref{2_eq_reform_1_2_1_4},\eqref{p_2},\eqref{p_3}, \notag \\
& \ \ \  \eqref{convex_1}-\eqref{convex_6}\notag,
\end{align} 
which can be directly solved by using the interior point method provided by CVX toolbox \cite{boyd2004convex}. The iteration process goes until it converges to a stationary point, which is proved to be one of the local minima of  \eqref{eq_primal_problem_2} \cite{full2022li}.

We assume  the obtained beamforming and rate-splitting solution is $\{\mathbf{c}^{(*)}, \mathbf{W}_c^{(+)},\mathbf{W}_k^{(+)} \}$. If $\mathrm{rank}(\mathbf{W}_c^{(+)})= 1$ and $\mathrm{rank}(\mathbf{W}_k^{(+)})= 1$, $ \forall k\in \mathcal{K}$, the relaxation is tight and the optimal solution $\mathbf{w}_c^{(*)},\mathbf{w}_k^{(*)}$ for the original problem \eqref{eq_primal_problem} can be extracted from $\mathbf{W}_c^{(+)},\mathbf{W}_k^{(+)}$ via eigenvalue decomposition, i.e., $\mathbf{W}^{(+)}= \mathbf{U} \Sigma \mathbf{U}^{\dag}$. Otherwise, $t$ is a upper bound of the minimum user rate. In this case, the beamforming solution can be scaled to the maximum length necessary to satisfy the constraints  according to Gaussian randomization method.  Moreover, according to \cite{li2018cooperative,mao2020maxmin,Polik2010Interior,Li2017joint}, the computation complexity of the proposed algorithm based on the interior point method is $\mathcal{O}(N^{3.5}\ \mathrm{log}(1/\epsilon))$, where $\epsilon$ and $N$ denote the given solution accuracy and number of decision variables, respectively. Assuming $I$ iterations are carried out before convergence, the overall computation complexity can be written as $I \cdot \mathcal{O}((KM+M+10K+1)^{3.5}\ \mathrm{log}(1/\epsilon))$.

For better illustration, the detailed procedure of this algorithm is summarized in Algorithm $\ref{Solution}$. 

\begin{algorithm} \label{Solution} 
            \caption{The Proposed Robust Iterative Algorithm   }
            1:  Initialize with a feasible point $\mathbf{z}^{(0)}=$ $\{ \beta_{c,k}^{(0)},x_{c,k}^{(0)}, y_{c,k}^{(0)}$,\\
        \ \ \     $\beta_{p,k}^{(0)}$, $x_{p,k}^{(0)}, y_{p,k}^{(0)}\}^T$, generated by the feasible point \\
        \ \ \ search algorithm  proposed in \ref{Feasible Point Search};  set  the iteration \\ 
         \ \ \ number $i$ = 0; define the tolerance of accuracy $\varepsilon$. \\
2: $\textbf{Repeat}$  \\
3:  \    Compute $\hat{f_1} ([x_{c,k}^{(i)}\ y_{c,k}^{(i)}]^T,[x_{c,k} \ y_{c,k}]^T)$, $\hat{f_2} (\beta_{c,k}^{(i)},\beta_{c,k})$, \\
\ \ \ \ $\hat{f_3} ( y_{c,k}^{(i)},y_{c,k})$, $\hat{f_4} (\beta_{p,k}^{(i)},\beta_{p,k})$ and  $\hat{f_5} ( y_{p,k}^{(i)},y_{p,k}) $ \\ 
\ \ \ \ according to \eqref{f_1}, \eqref{f_2},  \eqref{f_3}, \eqref{f_4}, \eqref{f_5} respectively. \\              
4: \    Solve the convex optimization problem $\eqref{eq_convex_problem}$, assign \\ \ \ \ \   the solution to  $\mathbf{z}^{(i+1)}$.\\
5: \  Update the iteration number:\ $i \leftarrow i+1 $. \\
6: $\textbf{Until}$  $|\{t (\mathbf{x}^{(i)}) \} - \{t (\mathbf{x}^{(i+1)}) \}| \leq $ $\varepsilon$; obtain the \\
 \ \ \ optimal solution $\{\mathbf{c}^{(*)}, \mathbf{W}_c^{(+)},\mathbf{W}_k^{(+)} \}$ to $\eqref{eq_convex_problem}$.\\
7: $\textbf{If}$  $\mathrm{rank}(\mathbf{W}_c^{(+)})= 1$ and $\mathrm{rank}(\mathbf{W}_k^{(+)})= 1$ \\
8: \   Extract the optimal solution $\mathbf{w}_c^{(*)},\mathbf{w}_k^{(*)}$   from $\mathbf{W}_c^{(+)}$,\\
 \ \ \   \ $\mathbf{W}_k^{(+)}$ via eigenvalue decomposition;\\
9: $\textbf{Else}$\\
10:   Adopt Gaussian randomization method  to generate a \\
\ \ \  \ \ feasible solution to \eqref{eq_rank_1_convex_problem}. 
\end{algorithm}

\subsection{Feasible Point Search Algorithm} \label{Feasible Point Search}
The generation of initial feasible point $\mathbf{z}^{(0)}$ in Algorithm $\ref{Solution}$ is critical and difficult to find.  In this section,  Algorithm $\ref{Feasible Point Search Algorithm}$ is proposed to produce an initial feasible point $\mathbf{z}^{(0)}$ and then to start Algorithm $\ref{Solution}$. 

In specific, we first set arbitrary initial values for $0 \leq c_k^{(0)} \leq \mathrm{r}_k, \  k\in \mathcal{K}$. Fixing the rate-splitting vector $\mathbf{c}^{(0)}$, the feasible transmitter beamformers $\mathbf{w}_{c}$ and  $\mathbf{w}_{k},\  k\in \mathcal{K}$  satisfying the  constraints \eqref{Init_constr_1}-\eqref{Init_constr_3}, can be found by solving the feasibility check
problem \eqref{Init_prob} of finding $\{ \mathbf{w}_{k} \},\mathbf{w}_{c}   $ , 
\begin{subequations} \label{Init_prob}
\begin{align} 
& \ \  \mathrm{Find}\ {\{ \mathbf{w}_{k} \},\mathbf{w}_{c} } \notag \\
  \text{s.t.} 
& \   \sum_{ j \in \mathcal{K}} c_j^{(0)}  \leq   R_{c,k}(\mathbf{w}_c,\mathbf{w}_k,\mathbf{h}_k ) ,\ \forall \mathbf{h}_k \in \mathcal{H}_k, \ \forall k \in \mathcal{K},  \label{Init_constr_1} \\
& \  0 \leq R_{p,k}(\mathbf{w}_k,\mathbf{h}_k ) ,\ \forall \mathbf{h}_k \in \mathcal{H}_k, \ \forall k\in \mathcal{K}. \label{Init_constr_2}\\
& \ \mathbf{w}_{c}^\dag \mathbf{w}_{c}+\sum_{k=1}^{K} \mathbf{w}_{k}^\dag \mathbf{w}_{k} \leq P_{\max}, \label{Init_constr_3}
\end{align}
\end{subequations}
To solve problem  \eqref{Init_prob}, we first reformulate it  into a  tractable form through the following proposition:

\begin{proposition} \label{prop_1}
The constraint \eqref{Init_constr_1} is equivalent to $ \mathbf{\xi}_{c,k}(\mathbf{w}_c,\mathbf{w}_k,\mathbf{h}_k )  \geq \mathrm{a}_{c,k} $, $\forall \mathbf{h}_k \in \mathcal{H}_k$, $\forall k\in \mathcal{K}$, the target SINR  $\mathrm{a}_{c,k}=e^{\sum_{ j \in \mathcal{K}} c_j^{(0)} +\frac{b_{c,k}^*}{2}-1}$, where $b_{c,k}^*=\mathcal{W}(^{2D_{c,k},-2D_{c,k}};-4e^{-2 \sum_{ j \in \mathcal{K}} c_j^{(0)} }D_{c,k}^2)$ with $\mathcal{W}(^{\tau_1,\tau_2};\mu)$ being given by
\begin{align} 
\mathcal{W}(^{\tau_1,\tau_2};\mu)& = \tau_1- \sum_{m=1}^{ \infty }\frac{1}{m*m!}\left( \frac{\mu me^{-\tau_1}}{\tau_2-\tau_1} \right)^m \notag \\ &  \sum_{n=1}^{ m-1 } \frac{(m-1+n)!}{n!(m-1-n)!} \left(\frac{-2}{2m(\tau_2-\tau_1)} \right)  ^n.
\end{align}
\end{proposition} 
\begin{IEEEproof}
Given $v_0 = \sqrt{\frac{1+\sqrt{1+4D_{c,k}^2}}{2}}-1$, $R(\gamma)$ with FBL is a monotonic decreasing function for $0 \leq \gamma \leq v_0$, where  $R(0)=0 $,  $R(v_0) \leq 0$ \cite{beamforming2021he}. And $R(\gamma)$  is a monotonic increasing function for $ \gamma \geq v_0$ \cite{beamforming2021he}. As the fixed common data rate  $\sum_{ j \in \mathcal{K}} c_j^{(0)}$ is no less than 0, i.e., $\sum_{ j \in \mathcal{K}} c_j^{(0)} \geq 0$,  we get that $R(\gamma)$ is a monotonic increasing function for $ \gamma \geq \mathrm{a}_{c,k}$ where  $R(\mathrm{a}_{c,k}) =\sum_{ j \in \mathcal{K}} c_j^{(0)}$. Consequently, 
\eqref{Init_constr_1} can be transformed  to $ \mathbf{\xi}_{c,k}(\mathbf{w}_c,\mathbf{h}_k )  \geq \mathrm{a}_{c,k} $, $\forall \mathbf{h}_k \in \mathcal{H}_k$, $\forall k\in \mathcal{K}$,  equivalently. 
\end{IEEEproof}

Through proposition \eqref{prop_1},  we can also obtain that \eqref{Init_constr_2} is equivalent to $ \mathbf{\xi}_{p,k}(\mathbf{w}_c,\mathbf{h}_k )  \geq \mathrm{a}_{p,k} $, $\forall \mathbf{h}_k \in \mathcal{H}_k$, $\forall k\in \mathcal{K}$, the target SINR  $\mathrm{a}_{p,k}=e^{ \frac{b_{p,k}^*}{2}-1}$, where $b_{p,k}^*=\mathcal{W}(^{2D_{p,k},-2D_{p,k}};-4D_{p,k}^2)$.

Based on the preliminaries above, the optimization problem  \eqref{Init_prob} can be  reformulated to 
\begin{subequations} \label{Initial_trans}
\begin{align} 
& \ \ \mathrm{Find}\ {\{ \mathbf{w}_{k} \},\mathbf{w}_{c} } \notag \\
  \text{s.t.} 
& \ \mathbf{\xi}_{c,k}(\mathbf{w}_c,\mathbf{w}_k,\mathbf{h}_k )  \geq \mathrm{a}_{c,k} ,\  \forall \mathbf{h}_k \in \mathcal{H}_k,\ \forall k\in \mathcal{K} \label{eq_robust_reform_2_2_1}\\
& \ \mathbf{\xi}_{p,k}(\mathbf{w}_k,\mathbf{h}_k )  \geq \mathrm{a}_{p,k} , \forall \mathbf{h}_k \in \mathcal{H}_k, \forall k\in \mathcal{K},\label{eq_robust_reform_2_2_2}\\
& \ \eqref{Init_constr_3}.\notag
\end{align}
\end{subequations}
 For the infinite constraints \eqref{eq_robust_reform_2_2_1} and \eqref{eq_robust_reform_2_2_2}, we again adopt  S-Procedure method to convert them into  LMI.  \eqref{eq_robust_reform_2_2_1} and \eqref{eq_robust_reform_2_2_2} can be converted into the following LMI constraints \cite{robust2014zhu},
\begin{equation}
\boldsymbol{\Xi}_{c,k}(\mathbf{W}_c,\mathbf{W}_k)\succeq\mathbf{0},\ k\in\mathcal{K},
\end{equation}
\begin{equation}
\boldsymbol{\Xi}_{p,k}(\mathbf{W}_k)\succeq\mathbf{0},\ k\in\mathcal{K},
\end{equation}
where 
\begin{align}
& \ \boldsymbol{\Xi}_{c,k}(\mathbf{W}_c,\mathbf{W}_k) = \notag \\
& \ \ \ \ 
\left[ 
\begin{array}{ccc}
\eta_{c,k} \mathbf{I}_M +\mathbf{U}_k & \mathbf{U}_k \hat{\mathbf{h}}_{k} \\
\hat{\mathbf{h}}_{k}^\dag \mathbf{U}_c ^\dag & \hat{\mathbf{h}}_{k}^\dag \mathbf{U}_k \hat{\mathbf{h}}_{k}-\eta_{c,k} \delta_k^2- \sigma_{k}^{2} \mathrm{a}_{c,k}   \\
\end{array} 
\right],\nonumber
\end{align}
\begin{align}
&\ \boldsymbol{\Xi}_{p,k}(\mathbf{W}_k)=\notag \\
& \ \ \ \ 
\left[ 
\begin{array}{ccc}
\eta_{p,k} \mathbf{I}_M + \mathbf{Q}_k & \mathrm{Q}_k \hat{\mathbf{h}}_{k} \\
\hat{\mathbf{h}}_{k}^\dag \mathbf{Q}_k ^\dag & \hat{\mathbf{h}}_{k}^\dag \mathbf{Q}_k \hat{\mathbf{h}}_{k}-\eta_{p,k} \delta_k^2- \sigma_{k}^{2} \mathrm{a}_{p,k}   \\
\end{array} 
\right].\nonumber
\end{align}
respectively, where $\mathbf{U}_k =\mathbf{W}_c- \sum_{j=1}^{K} \mathbf{W}_j$, $\mathbf{Q}_k =\mathbf{W}_k- \sum_{j=1,j \neq k}^{K} \mathbf{W}_j$.

Then the  problem  \eqref{Init_prob} can be  equivalently transformed into the  rank-one constrained  problem as
\begin{subequations} \label{eq_robust_2}
\begin{align}
& \ \ \ \mathrm{Find}\ {\{ \mathbf{w}_{k} \},\mathbf{w}_{c} } \notag \\
  \text{s.t.} 
 & \quad \boldsymbol{\Xi}_{c,k}(\mathbf{W}_c,\mathbf{W}_k)\succeq\mathbf{0},\ k\in\mathcal{K},  \\
 & \quad \boldsymbol{\Xi}_{p,k}(\mathbf{W}_k)\succeq\mathbf{0},\ k\in\mathcal{K},\\
  &  \quad \mathrm{rank}(\mathbf{W}_k)= 1,\ k\in\mathcal{K},\label{constr_2}\\
  & \quad \eqref{Init_constr_3}.\notag
\end{align}
\end{subequations}
We can adopt the Gaussian randomization method \cite{Robust2016liao} to generate a feasible solution to \eqref{Init_prob} or  the penalty function method \cite{phan2012nonsmooth} to provide a good rank-one approximation.

After obtaining the transmitter beamformers $\mathbf{w}_{c}^{(0)}$, $\mathbf{w}_k^{(0)}$ when $\mathbf{c}$ is fixed, $x_{c,k}^{(0)}$ is initialized by replacing the inequality of  $  e^{x_{c,k}} \leq | \mathbf{h}_{k}^\dag \mathbf{w}_{c}|^2, \ \forall \mathbf{h}_k \in \mathcal{H}_k,\ \forall k\in \mathcal{K}$
 with equality, i.e.,  
 \begin{align}
   e^{x_{c,k}^{(0)}} &= \min_{ \mathbf{h}_k \in \mathcal{H}_k} | \mathbf{h}_{k}^\dag \mathbf{w}_{c}^{(0)}|^2 \notag \\
   &= \min_{ \mathbf{h}_k \in \mathcal{H}_k} | (\hat{\mathbf{h}}_{k}^\dag+\Delta \mathbf{h}_{k}^\dag) \mathbf{w}_{c}^{(0)}|^2 \notag \\
   &= \min_{ \mathbf{h}_k \in \mathcal{H}_k} | \hat{\mathbf{h}}_{k}^\dag \mathbf{w}_{c}^{(0)}+\Delta \mathbf{h}_{k}^\dag \mathbf{w}_{c}^{(0)}|^2 \notag  \\
   &=(|\hat{\mathbf{h}}_{k}^\dag \mathbf{w}_{c}^{(0)}|-\delta_k||  \mathbf{w}_{c}^{(0)}||)^2, \ \forall k\in \mathcal{K}.
\end{align}

And $y_{c,k}^{(0)}$ is initialized by replacing the inequality of  $   e^{y_{c,k}} \geq  \sum_{j=1}^{K}| \mathbf{h}_{k}^\dag \mathbf{w}_{j} |^2+ \sigma_{k}^{2}, \ \forall \mathbf{h}_k \in \mathcal{H}_k$
 with equality, i.e.,  
 \begin{align}
   e^{y_{c,k}^{(0)}} &= \max_{ \mathbf{h}_k \in \mathcal{H}_k}  \sum_{j=1}^{K}| \mathbf{h}_{k}^\dag \mathbf{w}_{j} ^{(0)}|^2+ \sigma_{k}^{2} \notag \\
   &= \max_{ \mathbf{h}_k \in \mathcal{H}_k} \sum_{j=1}^{K}| (\hat{\mathbf{h}}_{k}^\dag+\Delta \mathbf{h}_{k}^\dag) \mathbf{w}_{j}^{(0)}|^2 + \sigma_{k}^{2}\notag \\
   &= \max_{ \mathbf{h}_k \in \mathcal{H}_k} \sum_{j=1}^{K} | \hat{\mathbf{h}}_{k}^\dag \mathbf{w}_{j}^{(0)}+\Delta \mathbf{h}_{k}^\dag \mathbf{w}_{j}^{(0)}|^2 + \sigma_{k}^{2} \notag  \\
   &=\sum_{j=1}^{K}(|\hat{\mathbf{h}}_{k}^\dag \mathbf{w}_{j}^{(0)}|+\delta_k||  \mathbf{w}_{j}^{(0)}||)^2 + \sigma_{k}^{2}, \ \forall k\in \mathcal{K}.
\end{align}

At last, $\beta_{c,k}$ can be initialized by replacing the inequality of  $  \beta_{c,k} \leq   e^{x_{c,k}-y_{c,k}} ,\ \forall k\in \mathcal{K}$  with equality, i.e.,  
 \begin{align}
  \beta_{c,k}^{(0)}= e^{x_{c,k}^{(0)}-y_{c,k}^{(0)}}, \ \forall k\in \mathcal{K}.
\end{align}

By using the same approach, $x_{p,k}^{(0)},y_{p,k}^{(0)},\beta_{p,k}^{(0)}$ can also be obtained.  For better illustration, the detailed initialization  procedure is summarized in Algorithm $\ref{Feasible Point Search Algorithm}$. In addition, the computation complexity of Algorithm $\ref{Feasible Point Search Algorithm}$ is related to the number of variables. It is worth noting that different  initial feasible point  may lead to different local minimum.

\begin{algorithm}\label{Feasible Point Search Algorithm}
\caption{Feasible Point Search Algorithm}
 1:  Generate $0 \leq c_k^{(0)} \leq \mathrm{r}_k, \  k\in \mathcal{K}$ randomly.\\
 2: Solve  problem \eqref{eq_robust_2}, and obtain the optimal solution \\
 \ \  \ $\{ \mathbf{W}_c^{(+)},\mathbf{W}_k^{(+)} \}$. \\
 3: $\textbf{If}$  $\mathrm{rank}(\mathbf{W}_c^{(+)})= 1$ and $\mathrm{rank}(\mathbf{W}_k^{(+)})= 1$ \\
4: \   Extract the optimal solution $\mathbf{w}_c^{(*)},\mathbf{w}_k^{(*)}$   from $\mathbf{W}_c^{(+)}$,\\
 \ \ \   \ $\mathbf{W}_k^{(+)}$ via eigenvalue decomposition;\\
5: $\textbf{Else}$\\
6:   Adopt Gaussian randomization method  to generate a \\
\ \ \   feasible solution $\{ \mathbf{w}_c^{(0)},\mathbf{w}_k^{(0)} \}$ to \eqref{Initial_trans}. \\
7:  Obtain the feasible
 $\mathbf{x}_{c}^{(0)} ,\mathbf{y}_{c}^{(0)},\boldsymbol{\beta}_{c}^{(0)} ,\mathbf{x}_{p}^{(0)} ,\mathbf{y}_{p}^{(0)},\boldsymbol{\beta}_p^{(0)} $: \\
\ \ \  \ ${x}_{c,k}^{(0)} =\mathrm{ln}(|\hat{\mathbf{h}}_{k}^\dag \mathbf{w}_{c}^{(0)}|-\delta_k||  \mathbf{w}_{c}^{(0)}||)^2, $      \\
\ \ \ \ ${y}_{c,k}^{(0)} =\mathrm{ln} [\sum_{j=1}^{K}(|\hat{\mathbf{h}}_{k}^\dag \mathbf{w}_{j}^{(0)}|+\delta_k||  \mathbf{w}_{j}^{(0)}||)^2+ \sigma_{k}^{2}],  $\\
\ \ \ \ $ \beta_{c,k}^{(0)}= e^{x_{c,k}^{(0)}-y_{c,k}^{(0)}}, $\\
\ \ \  \ ${x}_{p,k}^{(0)} =\mathrm{ln}(|\hat{\mathbf{h}}_{k}^\dag \mathbf{w}_{k}^{(0)}|-\delta_k||  \mathbf{w}_{k}^{(0)}||)^2, $  \\ 
\ \ \ \ ${y}_{p,k}^{(0)} =\mathrm{ln} [\sum_{j=1,j\neq k}^{K}(|\hat{\mathbf{h}}_{k}^\dag \mathbf{w}_{j}^{(0)}|+\delta_k||  \mathbf{w}_{j}^{(0)}||)^2+ \sigma_{k}^{2}],  $\\
\ \ \ \ $ \beta_{p,k}^{(0)}= e^{x_{p,k}^{(0)}-y_{p,k}^{(0)}},\ \forall k\in \mathcal{K} .$\\
\end{algorithm}

\section{Simulation Results and Analysis}\label{Simulation}

In this section, comprehensive simulations are carried out to evaluate the performance of our proposed robust  beamforming and rate-splitting design in the FBL regime for xURLLC, i.e., RB-RS-FBL. For comparison, the following  transmission schemes are  simulated and included as benchmarks  including:
\begin{itemize}
\item  NoRB-RS-FBL: It is a non-robust rate-splitting design in the FBL regime assuming perfect CSIT \cite{xu2021rate};
\item RB-NoRS-FBL: It is a robust design with no rate-splitting in the FBL regime assuming imperfect CSIT. As there is seldom  investigation on the robust beamforming design in the FBL regime, to compare, this scheme is obtained by simplifying our proposed RB-RS-FBL scheme  through setting $c_k =0$ for all $k \in \mathcal{K}$. 

\item RB-RS-IFBL: It is a robust rate-splitting design in the IFBL regime based on Shannon's  capacity assuming imperfect CSIT  \cite{joudeh2016robust}.
\end{itemize}

In  the following simulations, the system setups of all the schemes are the same. The  convergence performance  of the  proposed robust iterative algorithm  is  first evaluated, followed by the robustness performance of the   RB-RS-FBL design. Finally we  investigate the relations between the  max-min rate  and the transmission blocklength,  the BLER, the SNR and the CSIT quality, etc. 
In the simulations, the antenna-number of  the transmitter is set to $M=4$, and the noise power density density is $\sigma_k=\sigma=$ -20 dBm unless otherwise stated. Moreover, the tolerance of accuracy  $\varepsilon = 10^{-6}$.


\subsection{ Convergence Performance}
\label{Convergence}
\begin{figure}[!t]
  \centering
  \includegraphics[width=0.53\textwidth]{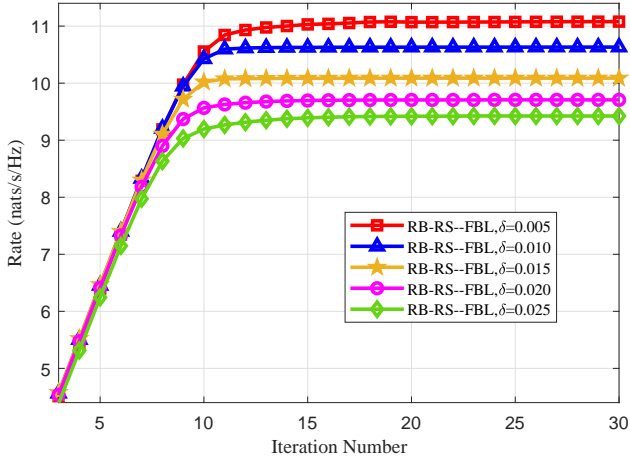}\\
  \caption{The convergence performance of the proposed iterative algorithm of the RB-RS-FBL design, where $M=4$, $K=2$, $L_{c,k}=L_{p,k}=L= 1000$, $\epsilon_k = \epsilon=10^{-5}$, $P_{\max}= 1000 $ mW.
  } \label{fig_convergence}
\end{figure}

First, we examine the  convergence performance of the proposed iterative  algorithm of the RB-RS-FBL design.  The total number of users is set to be $K=2$, the transmission block-length is set to be $L_{c,k}=L_{p,k}=L= 1000$,  the BLER is $\epsilon_k = \epsilon=10^{-5}$, $k \in \mathcal{K} $, and the maximum transmission power is $P_{\max}= 1000 $ mW.  All the channels  are assumed to follow both large-scale pass loss fading and small-scale Rayleigh fading. For simplicity, we adopt the random channel realizations in \cite{mao2018rate} by assuming that all the channels $\mathbf{h}_{k}$, $k \in \mathcal{K} $ follow independent and identically distributed (i.i.d.) complex Gaussian distribution, i.e., $\mathbf{h}_{k} \sim \mathcal{CN} (\mathbf{0},{\mathbf{I}_M})$.  And we assume that all the channel uncertainties are equally norm-bounded, i.e., $\delta_k = \delta$, $k \in \mathcal{K} $.

Under these  simulation setups, the convergence performances of the proposed algorithm for  non-scaling or fixed CSIT errors (i.e., $\alpha_k =\alpha= 0$)  are shown in Fig. \ref{fig_convergence} with  $\delta =0.005$, $\delta=0.010$, $\delta =0.015$, $\delta =0.020$  and $\delta =0.025$. The simulation results are averaged over 100 independent channel realizations. We observe that the  max-min rate for all the channel error cases converges to a stationary point after several iterations, which confirms the theoretical analysis. Moreover, the max-min rate of all the imperfect CSIT cases  decreases with the channel estimation error.

\subsection{ Robustness Performance }
\label{Robustness}
\begin{figure}[!t]
  \centering
  \includegraphics[width=0.52\textwidth]{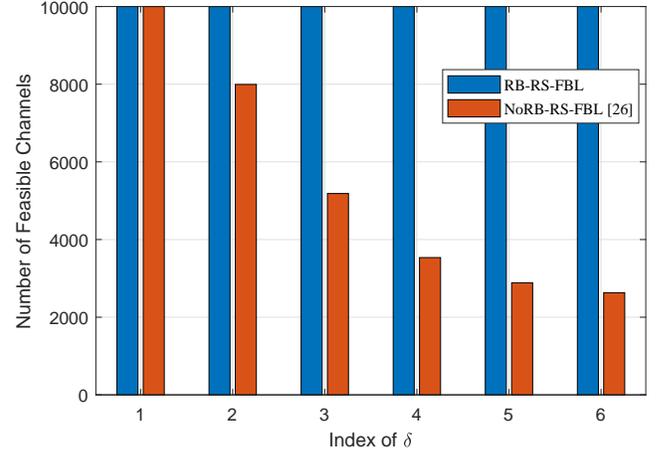}\\
  \caption{The number of feasible channel realizations for robust and non-robust algorithms under different channel estimation error, $\delta^2$ = $[  0 \ 10^{-12} \ 10^{-11} \  10^{-10} \  10^{-8} \  10^{-4} \ ]$, $M$ = $4$, $K$ = $2$,  $L$ = $1000$, $\epsilon = 10^{-5}$.
  }
   \label{fig_Robustness}
\end{figure}
To evaluate the robustness of the proposed  design, the robust rate-splitting algorithm (namely RB-RS-FBL)  and the non-robust rate-splitting algorithm (namely NoRB-RS-FBL) are simulated over 10000 channel realizations for  non-scaling CSIT errors ($\alpha_k =\alpha= 0$), i.e.,  $\delta_k^2 =\delta^2 = [ \ 0 \  10^{-12} \ 10^{-11} \  10^{-10} \  10^{-8} \  10^{-4} \ ]$, $k \in \mathcal{K}$. The non-robust algorithm adopts the  beamforming and rate-splitting design by assuming perfect channel acquisition. The total number of users is $K=2$, the blocklength is $L_{c,k}=L_{p,k}=L= 1000$,   the BLER is $\epsilon_k = \epsilon=10^{-5}$ for all $k \in \mathcal{K}$, and   the maximum transmission power is $P_{\max}= 1000 $ mW. As illustrated in Fig. \ref{fig_Robustness}, the RB-RS-FBL scheme yields feasible solution for all realizations. While the number of feasible channels for the non-robust algorithm NoRB-RS-FBL decreases as the CSIT uncertainty increases. This validates the robustness of the  RB-RS-FBL design  under imperfect CSIT cases.

\subsection{Max-Min  Rate Performance}

\subsubsection{Blocklength}
\label{Blocklength}

\begin{figure}[!t]
  \centering
 \includegraphics[width=0.53\textwidth]{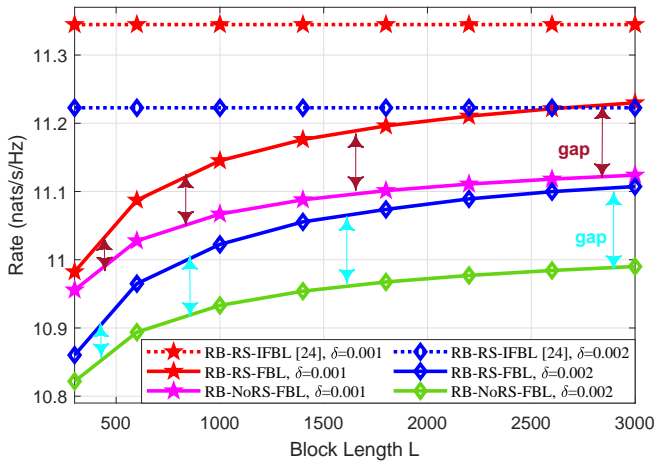}\\
  \caption{The max-min rate of the proposed RB-RS-FBL   and the baseline transmission schemes versus  the blocklength $L$ in the FBL and IFBL regime;  $\delta=0.001$, $\delta=0.002$; $M$ = $4$, $K$ = $2$,  $\epsilon = 10^{-5}$, $P_{\max}= 1000 $ mW.
    } \label{fig_Block_length}
\end{figure}

\begin{figure}[!t]
  \centering
 \includegraphics[width=0.53\textwidth]{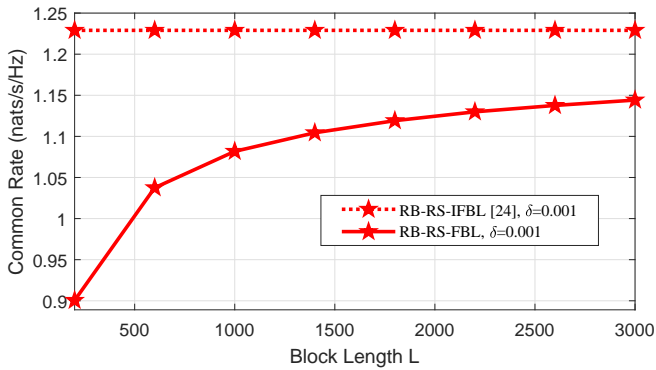}\\
  \caption{The sum common rate of the proposed RB-RS-FBL   in the FBL regime and the RB-RS-IFBL in the IFBL regime  versus  the blocklength $L$;  $\delta=0.001$; $M$ = $4$, $K$ = $2$,  $\epsilon = 10^{-5}$, $P_{\max}= 1000 $ mW.
  } \label{fig_common_rate}
\end{figure}

In this section, we investigate the relation between the robust max-min rate performance and the blocklength $L$. For comparison, the RB-RS-IFBL scheme \cite{joudeh2016robust} and the RB-NoRS-FBL scheme are also simulated and included. For the simulation setups, the total user number is $K=2$,  the BLER is $\epsilon_k = \epsilon=10^{-5}$ for all $k \in \mathcal{K}$, and the maximum transmission power is $P_{\max}= 1000 $ mW.  We adopt the specific channel realizations in \cite{mao2018rate,mao2018energy} by assuming that the channels are given by $\mathbf{h}_1 =[1,1,1,1]^{\dag}$ and $\mathbf{h}_2 = \gamma \ [1,e^{j\theta},e^{j2\theta},e^{j3\theta}]^{\dag}$, where $\gamma$ controls the channel strength disparity of the two users, and $\theta$ controls the channel correlation or channel angle disparity. For this simulation, we let $\gamma=0.9$, $\theta= 7 \pi/36 $.

By assuming the same system setups, the simulations of  RB-RS-FBL, RB-NoRS-FBL in the finite blocklength  regime [200,\ 3000],  and RB-RS-IFBL in the infinite blocklength  regime   are carried out using one specific channel realization for  non-scaling or fixed CSIT errors (i.e., $\alpha_k =\alpha= 0$)  with  $\delta =0.001$, $\delta=0.002$, respectively.  The obtained max-min rate with different transmission blocklength  is illustrated in Fig. \ref{fig_Block_length}.  In the figure, the two top straight dotted lines  are the max-min rate of RB-RS-IFBL for $\delta =0.001$ and $\delta=0.002$, respectively. Focusing on the two schemes in the FBL regime, i.e., RB-RS-FBL and RB-NoRS-FBL, we observe that the rate achieved by both of them increases with the blocklength $L$ and approaches that of RB-RS-IFBL when $L$ is getting large. This  verifies the fact that the low-latency transmission is at the expense of sacrificing the data rate. Moreover, RB-RS-FBL  outperforms RB-NoRS-FBL in all the blocklength regime, but when the blocklength $L $ is short, i.e., $L <500$, the performance gap between RB-RS-FBL and RB-NoRS-FBL is small. That is because in the FBL regime, the channel dispersion term $  \Lambda(\mathbf{\xi}_{k})= (1-(1+ \mathbf{\xi}_{c,k} )^{-2}) $ leads to  rate loss, and the common stream $s_c$ in RB-RS-FBL  will bring additional rate loss compared with RB-NoRS-FBL \cite{xu2021rate}. The shorter the blocklength, the greater  the rate loss.   It also can be verified through the following simulation.

To further investigate the rate-splitting process in the robust design, the sum common rate for all the users with different blocklength $L$ is shown in  Fig. \ref{fig_common_rate}. The  top  straight line is the common rate of  RB-RS-IFBL for $\delta =0.001$ when $L \rightarrow \infty$. We observe that the sum common rate of RB-RS-FBL in  CSIT error case increases with $L$. 
That is because RB-RS-FBL can balance the rate loss and the rate gain  by flexibly adjusting the rate-splitting ratio. That is, when $L$ is small,  the ratio of rate-splitting is adjusted to be  smaller to reduce the rate  loss. When $L$ increases, the common rate is increasing dramatically and approaching that of RB-RS-IFBL, as the rate gain brought by rate-splitting is much greater than the rate loss caused by additional common stream transmission.  The simulation result also  implies that it is better to decrease rate-splitting in extremely low latency transmissions.

\subsubsection{ BLER}
\label{BLER}
\begin{figure}[!t]
  \centering
  \includegraphics[width=0.53\textwidth]{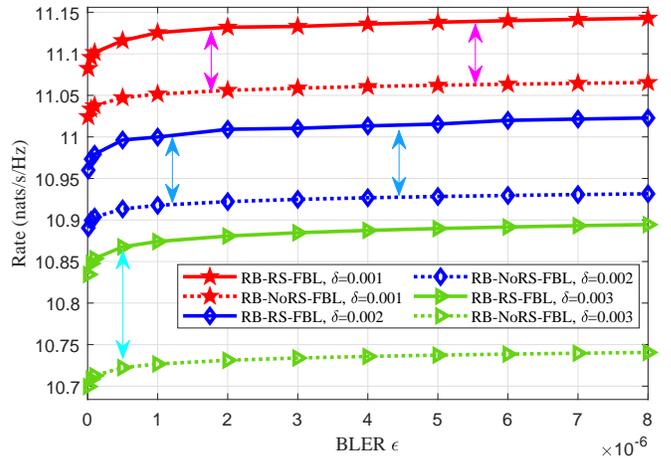}\\
  \caption{The max-min rate of the proposed RB-RS-FBL design  and the RB-NoRS-FBL   baseline  scheme   versus  BLER; $\delta=0.001$,  $\delta=0.002$, $\delta=0.003$;  $M$ = $4$, $K$ = $2$, $L$ = $1000$, $P_{\max}= 1000 $ mW.}
\label{fig_BLER}
\end{figure}

\begin{figure}[!t]
  \centering
  \includegraphics[width=0.53\textwidth]{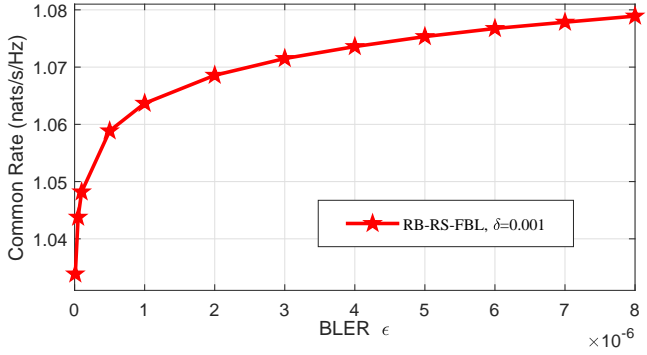}\\
  \caption{The common rate of the proposed RB-RS-FBL    in the FBL  regime versus  BLER for   $\delta=0.001$; $M$ = $4$, $K$ = $2$, $L$ = $1000$, $P_{\max}= 1000 $ mW.}
\label{fig_common_rate_BLER}
\end{figure}

To investigate the relation between the max-min rate performance and the BLER $\epsilon$, we assume that  $K=2$,  $L = 1000$, $P_{\max}= 1000 $ mW, and the channels  $\mathbf{h}_1 =[1,1,1,1]^{\dag}$ and $\mathbf{h}_2 = \gamma \ [1,e^{j \theta},e^{j2\theta},e^{j3\theta}]^{\dag}$, where $\gamma=0.9$, $\theta= 7 \pi/36 $. The simulations of  RB-NoRS-FBL  and the proposed RB-RS-FBL   with different BLER for  non-scaling CSIT errors with  $\delta =0.001$, $\delta=0.002$, $\delta =0.003$ are carried out using one specific channel realization. As shown in Fig. \ref{fig_BLER}, we observe that the obtained rate   of all the  schemes increases with the  BLER. It is worth noting that when the BLER is in a relatively small domain $\epsilon < 10^{-6}$,  the max-min rate varies dramatically. We conclude that high-reliability transmission is at the expense of sacrificing the  data rate. Moreover, the common rate  with different BLER is shown in  Fig. \ref{fig_common_rate_BLER}. It is found that the common rate of RB-RS-FBL increases with  BLER. This implies that it is better to decrease rate-splitting in ultra-high reliability transmissions.

\subsubsection{SNR}
\label{Scaling}

\begin{figure}[!t]
  \centering
 \includegraphics[width=0.53\textwidth]{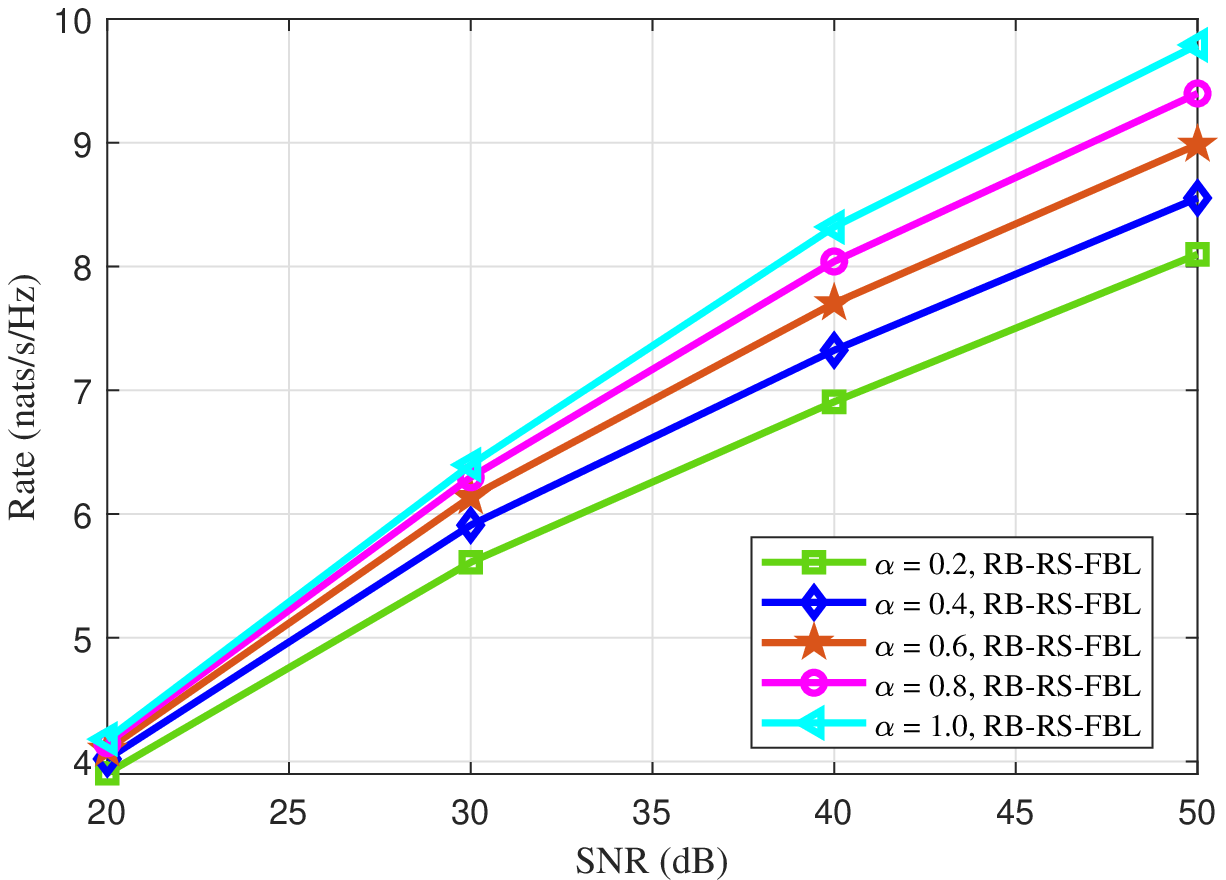}\\
  \caption{The max-min rate  of the proposed RB-RS-FBL design   for scaling CSIT errors versus the SNR with  $\alpha=0.2,\ 0.4,\ 0.6,\ 0.8, \ 1.0 $;  $M$ = $4$, $K$ = $2$, $L$ = $1000$, $\epsilon = 10^{-5}$.
  } \label{fig_SNR}
\end{figure}
\begin{figure}[!t]
  \centering
 \includegraphics[width=0.52\textwidth]{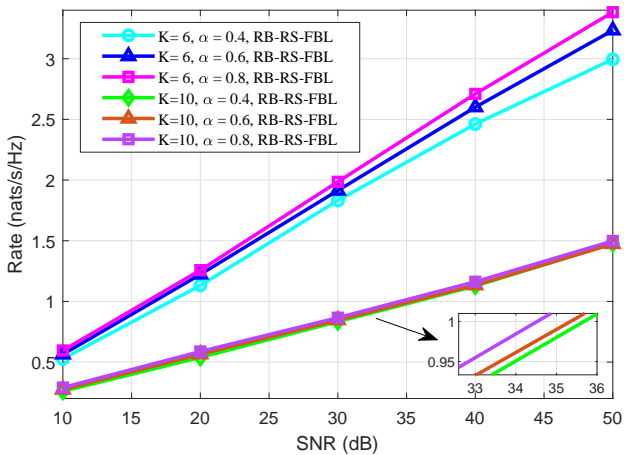}\\
  \caption{The max-min rate  of the proposed RB-RS-FBL design   for scaling CSIT errors versus the SNR with  $\alpha= 0.4,\ 0.6,\ 0.8 $;  $M$ = $4$, $L$ = $1000$, $\epsilon = 10^{-5}$.
  } \label{fig_UE}
\end{figure}

Finally, we evaluate the relation between the max-min rate performance and the SNR. The effect of CSIT quality on the minimum data rate is also evaluated for different number of users. For simplicity, the noise variance is  set to be 1, i.e., $\sigma_{k}=\sigma= 1$, then the transmission power $P_t = \mathrm{SNR}$. The
error variance is given as $\delta_k = \delta = d P_t^{-\alpha}$, where $d$  represents different CSIT accuracies. Moreover, we set $\epsilon_k = \epsilon=10^{-5}$,   $L = 1000$, $K=2$ and adopt the random channel realizations  by assuming  $\mathbf{h}_{k} \sim \mathcal{CN} (\mathbf{0},{\mathbf{I}_M})$. The simulation results are averaged over 100 independent channel realizations.

The obtained max-min rate  of the proposed RB-RS-FBL design  for   $\alpha=0.2$, $\alpha=0.4$, $\alpha=0.6$, $\alpha=0.8$ and $\alpha=1.0$ are illustrated in Fig. \ref{fig_SNR}. From the  figure we observe that the rate in all the CSIT error cases increases with SNR, and also increase with the CSIT quality $\alpha$. The gap between them grows larger with increased SNR. This implied that increasing the number of feedback bits when  SNR is in high regime  can improve the xURLLC system performance. For $K=6$ and $K=10$, the obtained rate  of the proposed RB-RS-FBL  for   $\alpha=0.4$, $\alpha=0.6$ and $\alpha=0.8$ are illustrated in Fig. \ref{fig_UE}. Obviously, the more the  users in the xURLLC system, the smaller the max-min data rate.

\section{Conclusion}
\label{conclusion}
In this work, we have proposed a robust  beamforming and rate-splitting design to enhance the robustness and improve the effectiveness  of  the xURLLC systems for  multi-user multi-antenna systems under imperfect CSIT cases.  We formulated the max-min rate problem by optimizing the beamforming vectors  and the  rate-splitting vector, under the premise of ensuring the requirements of transmission latency and transmission reliability. To solve the formulated non-convex optimization problem, the efficient  iterative algorithm based on  the CCCP and the Gaussian randomization method was proposed to obtain a local minimum after the transformation of the original problem and the transformation into  a  DC programming one. To start the iterative algorithm, we further proposed  an  intuitive initial feasible point search algorithm   by making utilization of  the convexity and concavity of the asymptotic capacity  in the FBL regime and by adopting the Gaussian randomization method. Simulation results confirmed the convergence, the robustness and the effectiveness of the proposed iterative algorithm, and revealed  that the robust design outperforms the other transmission schemes  under various  blocklength, BLER. Moreover, we concluded that robust  low-latency and high-reliability transmission are  at the expense of sacrificing the system effectiveness such as data rate.

\appendices

\bibliographystyle{IEEEtran} 
\bibliography{IEEEabrv,bib}

\end{document}